\documentclass{jaa}
\usepackage{natbib}
\usepackage{gensymb}
\usepackage{hyperref}
\usepackage{caption}
\usepackage{pdfpages}
\usepackage{float}
\usepackage{multicol}
\usepackage{dblfloatfix}
\usepackage{graphicx}
\usepackage{subcaption}
\usepackage[export]{adjustbox} % For valign adjustment if needed

\urldef\myurl\url{https://irsa.ipac.caltech.edu/cgi-bin/Gator/nph-scan?projshort=ZTF&mission=irsa}


\begin{document}\sloppy

%%paper title
%%For line breaks \\ can be used within title
\title{Performance of the 4-m International Liquid Mirror Telescope tested in two fields at high and low ecliptic and galactic latitudes}

%%author names are separated by comma (,)
%%use \and before the last author name
%%use a * along with the number separated by comma
%% for the  author for correspondence
%%\textsuperscript{number} is used for affiliation
%%\affilOne, \affilTwo etc., upto \affilTwentyfive is possible
%%Please note the first letter after \affil is capitalised in the command
%%

%\author{AUTHOR1\textsuperscript{1}, AUTHOR2\textsuperscript{1} and AUTHOR3\textsuperscript{2,*}}

\author{Sara Filali\textsuperscript{1,*},
         Kumar Pranshu\textsuperscript{2,3},
         Jean Surdej\textsuperscript{1,2},
         Paul Hickson\textsuperscript{4,5},
         Kuntal Misra\textsuperscript{2},
         Bhavya Ailawadhi\textsuperscript{2,6},
         Talat Akhunov\textsuperscript{7,8},
         Monalisa Dubey\textsuperscript{2,9},
         Naveen Dukiya\textsuperscript{2,9},
         Brajesh Kumar\textsuperscript{10},
         Priyanshi Kumari\textsuperscript{2},
         Vibhore Negi\textsuperscript{11},
         and
         Anna Pospieszalska-Surdej\textsuperscript{1}
         }

\affilOne{\textsuperscript{1}Institute of Astrophysics and Geophysics, Liège University, Allée du 6 août 19c, 4000 Liège, Belgium.\\}
\affilTwo{\textsuperscript{2}Aryabhatta Research Institute of Observational Sciences, Manora Peak, Nainital, Uttarakhand, 263001, India.\\}
\affilThree{\textsuperscript{3}University of Calcutta, 87/1 College Street, Kolkata, 700073, India.\\}
\affilFour{\textsuperscript{4}Department of Physics and Astronomy, The University of British Columbia, 6224 Agricultural Road, Vancouver, BC, V6T 1Z1, Canada.\\}
\affilFive{\textsuperscript{5}Outer Space Institute, The University of British Columbia, 325-6224 Agricultural Rd., Vancouver, BC, V6T 1Z1.\\}
\affilSix{\textsuperscript{6}Deen Dayal Upadhyay Gorakhpur University, Civil Lines, Gorakhpur, Uttar Pradesh, 273009, India.\\}
\affilSeven{\textsuperscript{7}National University of Uzbekistan, Department of Astronomy and Astrophysics, 100174 Tashkent, Uzbekistan.\\}
\affilEight{\textsuperscript{8}Ulugh Beg Astronomical Institute of the Uzbek Academy of Sciences, Astronomicheskaya 33, 100052 Tashkent, Uzbekistan.\\}
\affilNine{\textsuperscript{9}Mahatma Jyotiba Phule Rohilkhand University, Pilibhit Bypass Road, Bareilly, Uttar Pradesh, 243006, India.\\}
\affilTen{\textsuperscript{10}South-Western Institute for Astronomy Research, Yunnan University, Kunming 650500, Yunnan, P. R. China.\\}
\affilEleven{\textsuperscript{11}Inter University Center for Astronomy and Astrophysics, Post Bag 4, Ganeshkhind, Pune 411007, India.\\}

%%escape two column mode for title, affiliation and abstract
%%by giving \twocolumn command as shown

\twocolumn[{

\maketitle

%%include \corres to print the corresponding author Email id
\corres{sfilali@uliege.be}

%%include \msinfo for
%%manuscript information such as
%%received, revised and accepted dates
%%
%\msinfo{18 February 2025}{18 February 2025}

%%abstract
\begin{abstract}
The 4-m International Liquid Mirror Telescope (ILMT) offers a unique opportunity to detect transients in a narrow strip of sky. We explore ILMT's potential to detect astrometric and photometric transients at various ecliptic and galactic latitudes.
We inspected CCD frames observed at both low and high ecliptic and galactic latitudes during the commissioning phase and the November 2023 - May 2024 observation cycle, respectively.
We analysed these images using both visual inspection and the ILMT's transient detection and candidate classification pipeline. In the low ecliptic and galactic latitude field, we detected more than 500 transient candidates. We cross-matched these with the Minor Planet Checker (MPC) database, identifying 504 catalogued asteroids, all with predicted V-magnitudes brighter than 24 mag, representing a total of 152 distinct asteroids. We performed the same steps on the high ecliptic and galactic latitude field, detecting 30 MPC-catalogued asteroids, and one newly discovered photometric transient, named AT 2024fxn.
We present the positions, trajectories, and magnitudes of the detected asteroids observed in the SDSS \textit{g'}, \textit{r'}, and \textit{i'} spectral bands and compare results from both fields. We explore the lightcurve of AT 2024fxn, which shows partial compatibility with a supernova (SN) hypothesis, while the data invites further insights.
\end{abstract}

%%insert keywords separated by 3 hyphens using \keywords{words}

}]

%%close the twocolumn escape here

%%include \doinum{number}for the DOI number in the header
%%include \volnum{number} for the volume number in the header
%%include \year{yyyy} for  year of publication in the header
%%include \pgrange{num--num} page range of article in the header
%%include \artcitid{num} for the article citation id
%%include \lp to print last page of the article
%%include \setcounter{page}{pagenum} for the exact starting page of the article

\doinum{12.3456/s78910-011-012-3}
\artcitid{\#\#\#\#}
\volnum{000}
\year{0000}
\pgrange{1--}
\setcounter{page}{1}
\lp{1}

\section{Introduction}
   The International Liquid Mirror Telescope (ILMT) is a telescope dedicated to photometric and astrometric survey, installed at the Devasthal observatory in the foothills of the Himalayas at $2378$ m altitude and ($29\degree 21'41.4''$, $79\degree 41' 07.08''$) latitude and longitude. The telescope is permanently oriented toward the zenith, enabling it to continuously scan the same strip of the sky centred around the declination corresponding to the latitude of the site. As a result, its cadence is approximately one day under optimal conditions, except when affected by meteorological conditions. It is equipped with a 4096×4096 pixel CCD camera that operates in the Time Delay Integration (TDI) mode (\citealp{gibson1992time, surdej20234mb, surdej20254}), with an integration time $t_{int} = 102.35 \ s$. The CCD camera has a field of view of $22.3'$ x $22.3'$ and the signal integration operates along the right ascension ($\alpha$). The acquisition time of a single long CCD/TDI frame is approximately 17-min, corresponding to tenfold the integration time. Each portion of the strip observed over this timeframe, excluding the first ramping $22.3'$ x $22.3'$ image, constitutes an individual frame of the ILMT, having a width of $22.3'$ along the declination (Dec) and a length of $200.7'$ along $\alpha$. The high cadence of the telescope, its large diameter and small focal ratio ($f/D \approx 2.4$), make it a unique tool to detect minor planets (for more details on the ILMT, refer to \citealt{surdej20184}, \citeyear{surdej20234mb}).
   
   The ILMT has completed four observing cycles at the time of writing this paper: April–May and October–November 2022, March–June 2023, November 2023–May 2024, and October 2024–June 2025. At the beginning of the commissioning phase in October-November 2022, we recorded 20 contiguous CCD frames with the \textit{g'}, \textit{r'}, and \textit{i'} SDSS spectral filters. From these, we selected a 1.24 square-degree field corresponding to a single CCD footprint, observed repeatedly over nine nights and in three different filters. The field starts at $\alpha_{2000} = $ 4h 50m, corresponding to low ecliptic and galactic latitudes ( $6^\circ\lesssim \beta \lesssim 7^\circ$, $-10^\circ \lesssim b \lesssim -7^\circ$). We refer to this region as Field 0450 to reflect its right ascension.
   
   We subsequently analysed images acquired during February–May 2024, concentrating on a 6.22 square-degree field starting at $\alpha_{2000} = $ 13h 17m and corresponding to high ecliptic and galactic latitudes ($35^\circ\lesssim \beta \lesssim 43^\circ$, $66^\circ\lesssim b \lesssim  82^\circ$), which we refer to as Field 1317. In contrast to Field 0450 described earlier, this larger angular area was selected to ensure a sufficient coverage of transients, thereby enabling a robust statistical analysis. Our goal was to assess the potential of the ILMT observations at different ecliptic and galactic latitudes.\\
   The paper is structured as follows. In Section \ref{sec:asteroids}, we present the methodology, results, and analysis of asteroid detections in both fields. In Section \ref{sec:phot_transient}, we present the methodology, results and analysis of the detected photometric transient AT 2024fxn. In Section \ref{sec:conclusions}, we draw conclusions regarding the overall detections. In \protect\ref{detection_tables}, we provide the detected asteroid tables. Details on these tables can be found in \ref{appendix_classes} and \ref{notes_tables}. \ref{appen:phot_params} describes the photometric parameters used for the asteroid measurements, with the reported uncertainties limited to zero-point variations, while the full photometric uncertainty budget is described and discussed in Section \ref{subsec:mag_and_pos}.
%__________________________________________________________________

\section{Asteroids}
\label{sec:asteroids}
\subsection{Detection methodology}

\begin{figure}
    \resizebox{\hsize}{!}{\includegraphics{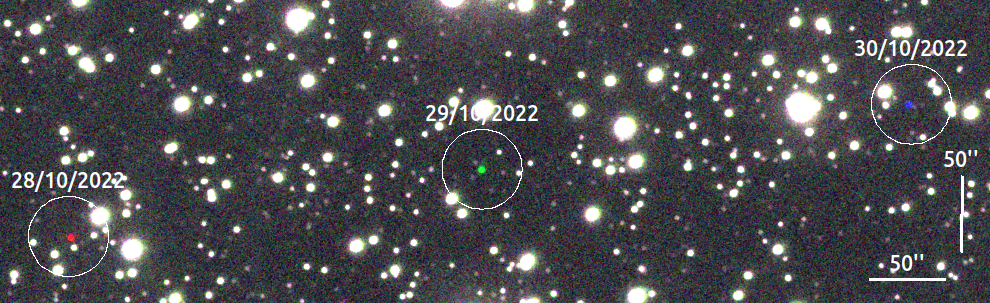}}
    \caption{Visual detection through RGB stacking. Red, green, and blue represent detections of the asteroid Melanthios on the nights of 28, 29, and 30th of October 2022, respectively, all obtained with the same \textit{i'}-SDSS filter. North is up, and East is to the left. The RGB method highlights temporal changes by assigning a distinct color component to each nighttime frame.}
    \label{fig:dates_melanthios_i2}
\end{figure}
\begin{figure*}
  \includegraphics[width=17cm, height=.18\textheight]{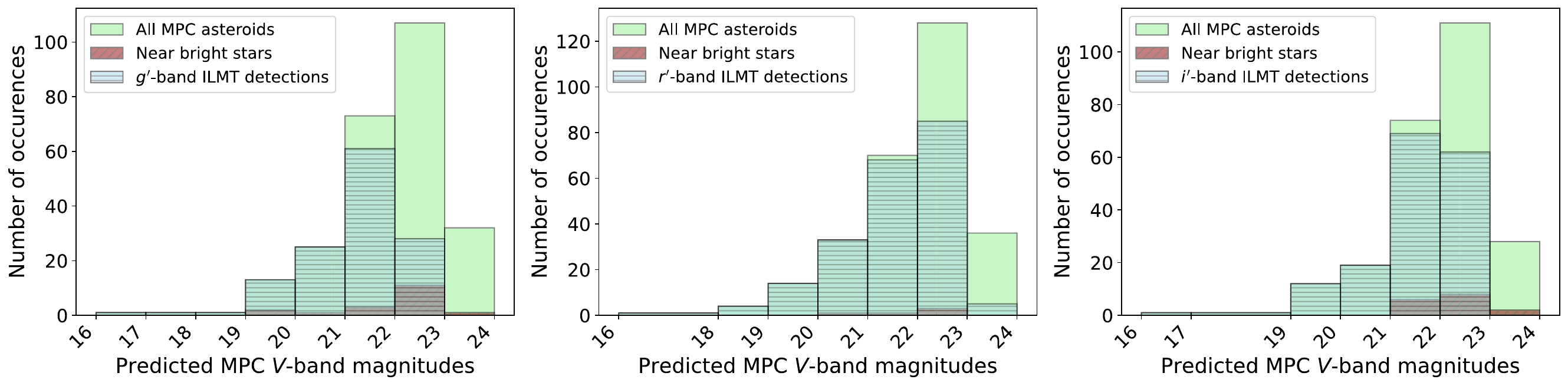}
  \caption{Distribution of the MPC-predicted \textit{V}-band magnitudes of asteroids detected in Field 0450 through the \textit{g'}, \textit{r'}, and \textit{i'} SDSS spectral filters.}
  \label{fig:dist_vmag_0450}
\end{figure*}
\begin{figure*}
\centering
  \includegraphics[width=10cm, height=.20\textheight]{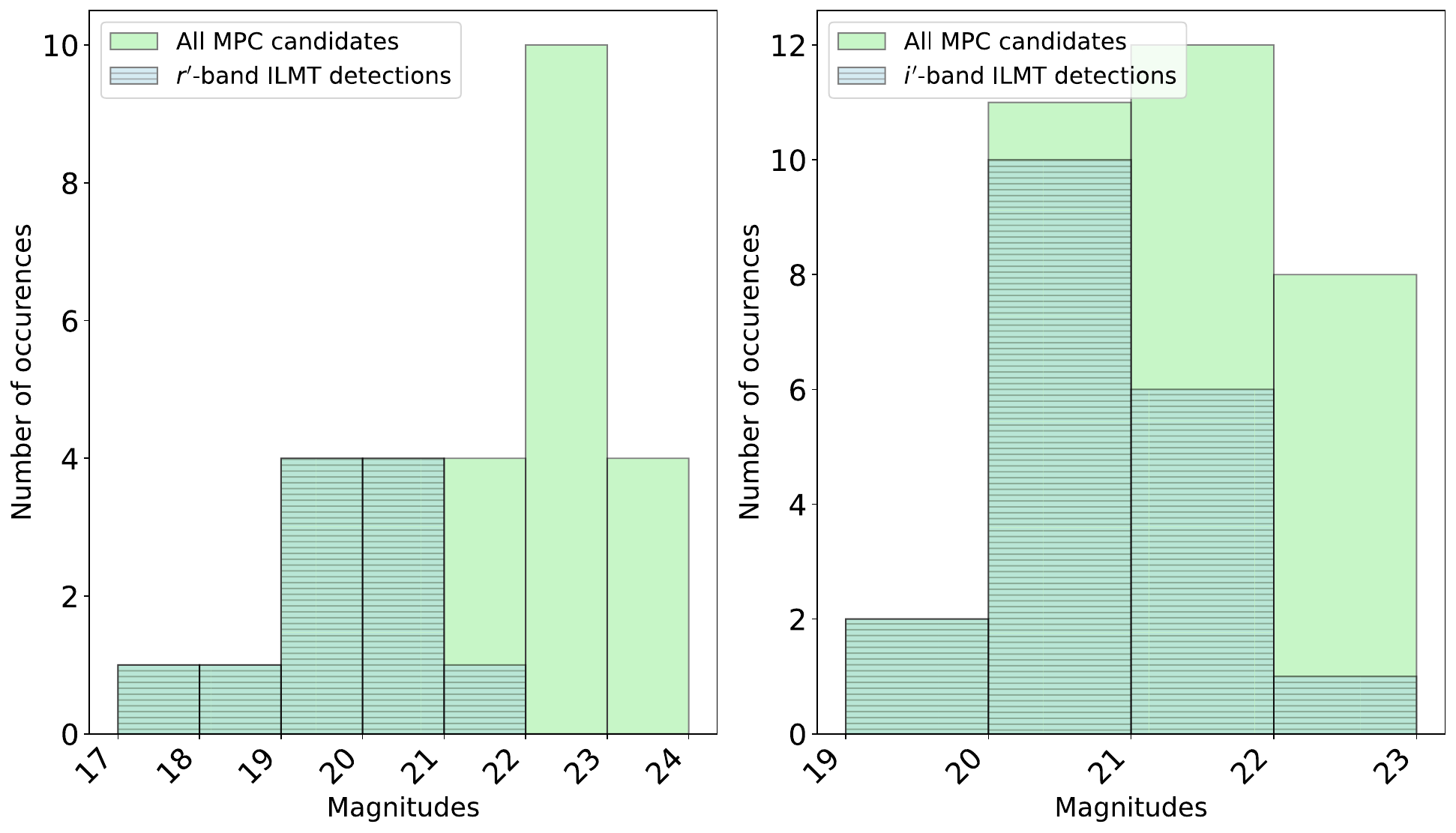} % smaller width so it looks less streteched
  \caption{Distribution of the MPC-predicted \textit{V}-band magnitudes of asteroids detected in Field 1317 through the \textit{r'} and \textit{i'} SDSS spectral filters.}
  \label{fig:dist_vmag_1400}
\end{figure*}
In Field 0450, we examined 9 CCD frames visually as well as with the PyLMT automated transient detection and candidate classification pipeline \citep{pranshu2025pylmt}. For the visual inspection, we employed the DS9 software \citep{joye2003new}, applying the blinking method to some frames, and the RGB stacking method to others (cf. \citealt{pospieszalska2023detection}). In the RGB stacking method, we aligned three frames captured on different nights, assigning each frame to one of the RGB color channels. In the resulting image, most sources appear white, while astrometric transients stand out in different colors depending on the night of observation. Figure \ref{fig:dates_melanthios_i2} shows an example of visual detection through RGB stacking in Field 0450. The PyLMT pipeline combines a subtraction technique module, a transient search and candidate classification module utilising a convolutional neural network (CNN), and a candidate cross-checking module \citep{pranshu2025pylmt}.
Applying both the visual and the PyLMT methods independently, we pinpointed several hundred transient candidates in total, with the PyLMT alone detecting 496 candidates. We cross-matched both methods and incorporated the exclusively visual detections in the PyLMT software to refine its detection parameters. Subsequently, we compared the combined detections from both approaches against the Minor Planet Center (MPC) database, and identified 504 detections of known asteroids within this field, corresponding to 152 distinct asteroids. These objects display predicted \textit{V}-magnitudes ranging between 16 and 24 mag. Although our limiting magnitude is estimated at $V\approx22$, we retained $V=24$ mag as the threshold to include a representative set of faint MPC-predicted asteroids for completeness assessment, while keeping the sample within a manageable range. This choice also means that the reported detection percentages are conservative; restricting the MPC search to $V \leq 22$ \citep{surdej20254} would have increased these rates.
Detection rates are: 130 detections out of 271 MPC predictions in the \textit{g'} band ($\sim$$48\%$ detection rate), 210 out of 286 in the \textit{r'} band ($\sim$$73.4\%$), and 164 out of 257 ($\sim$$63.8\%$) in the \textit{i'} band. Details of these detections are presented in \ref{detection_tables}, and their \textit{V}-magnitude distributions are presented in Figure \ref{fig:dist_vmag_0450}.

In Field 1317, we analysed 19 CCD frames observed with the ILMT in the \textit{r'} and \textit{i'} bands. The MPC predicted 61 asteroids, and we applied the now finer-tuned PyLMT pipeline for independent detection and classification, where image subtraction has been executed using frames from the same cycle as references. We performed a subsequent visual examination solely for confirmation. In total, 30 asteroids were detected by the PyLMT, out of which 22 are distinct ones. Detection rates are 11 out of 28 MPC predictions ($\sim$39.3 $\%$) in the \textit{r'} band, and 19 out of 33 ($\sim$57.6$\%$) in the \textit{i'} band. Differences in detectability with Field 0450 are explained by variations in the sky quality during observations. Additionally, the presence of more fainter asteroids in Field 1317, particularly in the 23–24 magnitude range, may have contributed to this effect. The \textit{V}-magnitude distributions of these detections are presented in Figure \ref{fig:dist_vmag_1400}. Defining the limiting recall magnitude as the magnitude at which half the asteroids are detected, Figure \ref{fig:dist_vmag_0450} shows that the ILMT limiting recall \textit{V}-magnitudes using the PyLMT pipeline in Field 0450 are 21.5, 22.0, and 22.0 in the \textit{g'}, \textit{r'}, and \textit{i'} bands, respectively (conservative estimates). For Field 1317, Figure \ref{fig:dist_vmag_1400} indicates that the limiting recall magnitude is 21 in both the \textit{r'} and \textit{i'} bands, also based on conservative estimates.
%__________________________________________________________________

\subsection{Magnitude and position measurements}
\label{subsec:mag_and_pos}
The typical point spread function (PSF) in our images has a mean full width at half maximum (FWHM) of approximately 1.5 arcsec and is close to circular across all filters, with median ellipticities $e\lesssim0.04$ in the \textit{r'} and \textit{i'} bands and $e\simeq0.12$ in the \textit{g'} band. The slightly higher ellipticity measured in the \textit{g'} band is attributed to  the less optimal corrections of the optical corrector (optimised in the spectral range $\lambda \simeq 5500-7500 \AA$) and the lower signal-to-noise ratio (S/N) of the \textit{g'} exposures, rather than to systematic tracking-induced elongation, indicating that the exposures themselves do not introduce significant elongation.
This is a result of the 5-lens asymmetric optical corrector \citep{hickson1998curvature} installed at the prime focus of the ILMT, which compensates for the nonlinear motion of stars in the images and largely eliminates TDI distortions.

After examining the apparent sky-plane rates of motion provided by MPChecker for asteroids in Fields 0450 and 1317, we found that the median asteroid motions in both fields are below $35$ arcsec/hr. For the exposure time of 102.35 s, this corresponds to a displacement of approximately 1 arcsec, which is smaller than the measured stellar PSF FWHM. Therefore, any elongation due to asteroid motion is small relative to the PSF, and asteroids can be treated as effectively point-like for the purposes of aperture photometry.
Thus, there is no need to consider a possible elongation of the PSF for the detected asteroids, and circular photometric apertures were used throughout the analysis. For the small number of sources exhibiting visually elongated or poorly defined centroid profiles (due to low S/N or unusually high motions), the corresponding photometric measurements were flagged as uncertain and treated accordingly in the subsequent analysis.
We employed two distinct methods for estimating magnitude and position, depending on the location of the target asteroid within the CCD frame:

\begin{itemize}
\item\textit{The statistical method:}
This method involves a statistical approach. The photometric zero point is estimated for each long CCD frame using Gaia reference stars present in it. The FWHM of the PSF is also estimated for each frame, rather than for individual objects \citep{2019hicksonLPICo2109.6066H}. The positions are estimated by fitting a centroid profile to each PSF, and aperture photomery is then applied with a radius of $2\times FWHM$. If the studied asteroid is located in a crowded region or in a region with high background gradient, we measured its magnitude using the \textit{individualised} method.

\item\textit{The individualised method:}
Aperture photometry is applied on the given asteroid using the imexam library \citep{sosey2017imexam}, which is affiliated to the Astropy project \citep{price2018astropy}.
The zero point for each asteroid is estimated by utilising neighbouring reference stars, with their number usually ranging between 3 and 7, and occasionally equal to 2. These stars are selected to minimise the root mean square (RMS) deviations of the magnitude zero points and the angular distances to the target asteroid. The asteroid's position is obtained by fitting a Gaussian profile to its recorded image.
\end{itemize}

\begin{figure*}
    \includegraphics[width=17cm, height=.25\textheight]{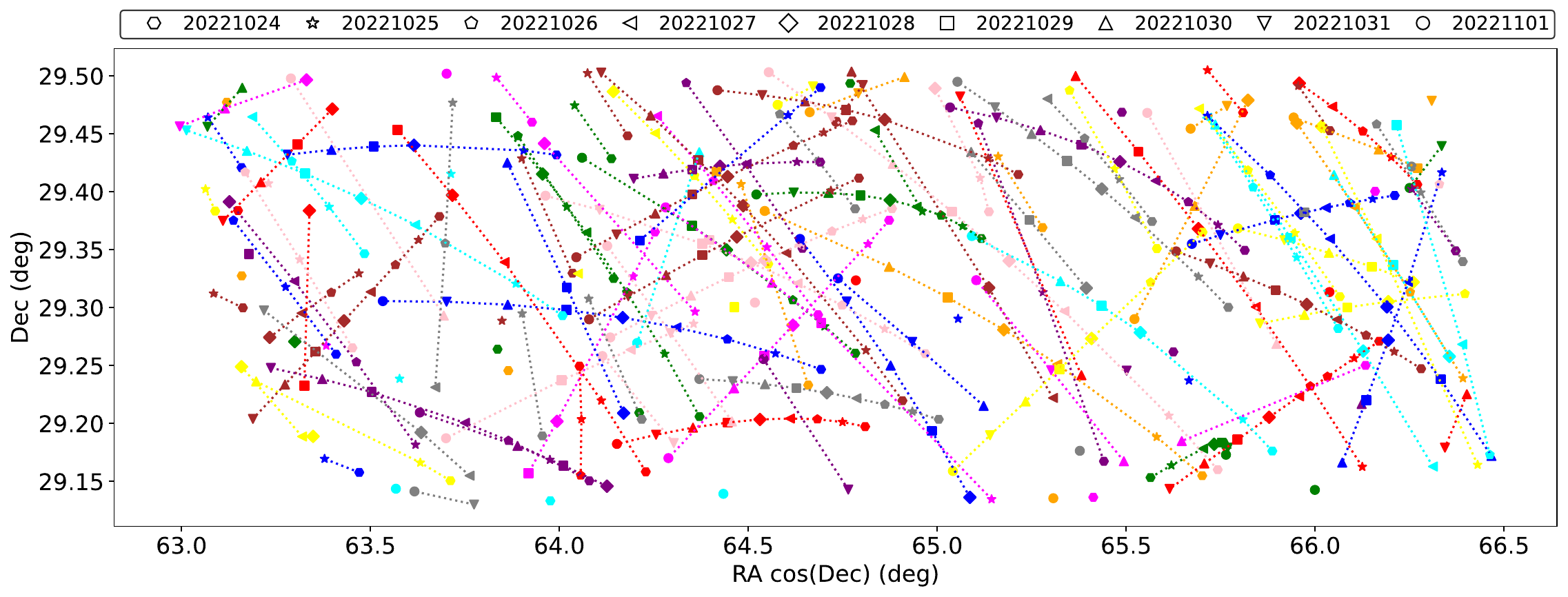}
    \caption{Positions and paths of the 504 asteroids detected with the ILMT in Field 0450 during the 24/10/2022 - 01/11/2022 nights. A unique color was assigned to each asteroid by cycling through a fixed color palette. Although some colors may correspond to more than one object, each asteroid is consistently represented by a single color and path across the figure. Three of the detected asteroids are confirmed to exhibit prograde motion, while 108 are confirmed to exhibit retrograde motion.}
    \label{fig:paths_0450}
    \vspace{1cm}
    \includegraphics[width=17cm, height=.25\textheight]{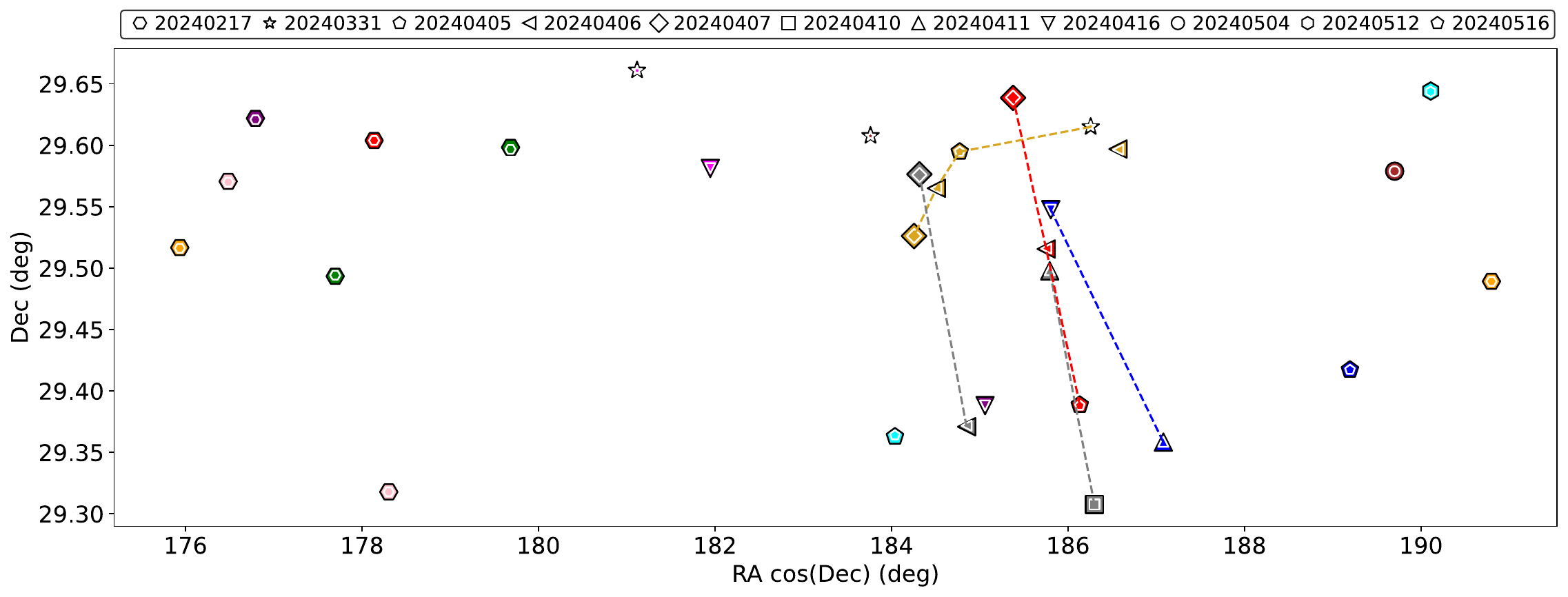}
    \caption{Positions and paths of the 30 asteroids detected with the ILMT in Field 1317 during the 14/02/2024 to 16/05/2024 nights. Observed and MPC-predicted positions share the same color per asteroid, though colors may repeat across different asteroids. Observed positions have larger markers with black edges; predicted positions have smaller markers with white edges, so predicted points appear inside observed ones---illustrating the close angular separations. Observed paths are shown with dashed lines. None exhibits prograde motion, and five are confirmed to exhibit retrograde motion. As expected, due to the higher ecliptic latitude of Field 1317, far fewer asteroids are detected compared to the Field 0450.}
    \label{fig:paths_1317}
\end{figure*}

\begin{figure*}
\centering
  \includegraphics[width=17cm]{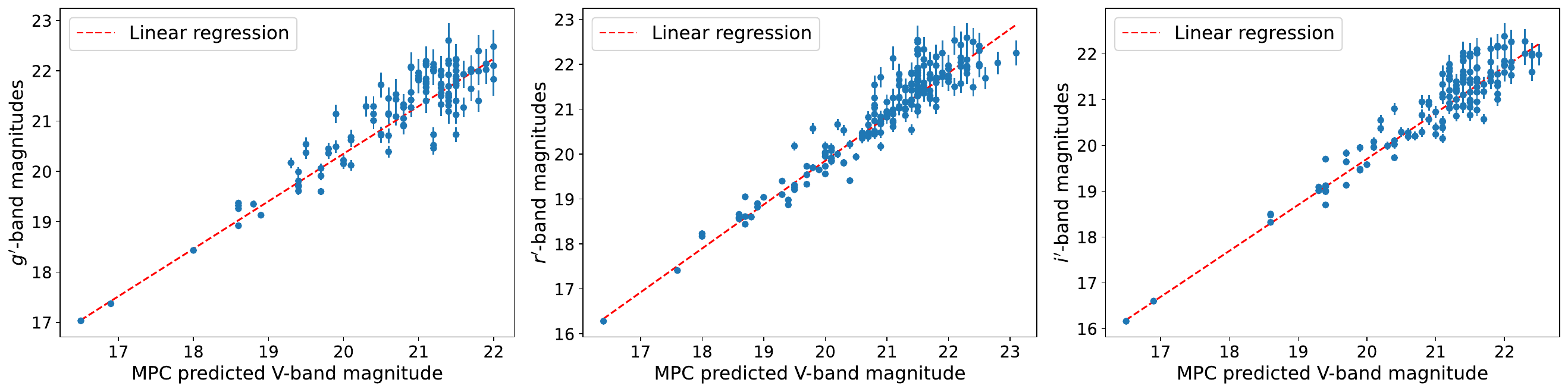}
  \caption{Measured SDSS-band magnitudes versus MPC-predicted \textit{V}-band magnitudes for the detected asteroids in Field 0450.}
  \label{fig:mag_meas_vs_mpc}
\end{figure*}

Figures \ref{fig:paths_0450} and \ref{fig:paths_1317} display the paths of asteroids as determined from position measurements within their respective fields.
Furthermore, we plotted the \textit{g'}, \textit{r'}, and \textit{i'} measured magnitudes versus the MPC-predicted \textit{V}-magnitudes in Field 0450 and applied linear regression, as shown in Figure \ref{fig:mag_meas_vs_mpc}. The obtained relations are:
\begin{align}
\textit{g'} &= (0.94 \pm 0.01)\, \textit{V} + (1.46 \pm 0.11) \label{eq:gp_vs_V} \\
\textit{r'} &= (0.98 \pm 0.01)\, \textit{V} + (0.31 \pm 0.08) \label{eq:rp_vs_V} \\
\textit{i'} &= (1.00 \pm 0.01)\, \textit{V} - (0.38 \pm 0.07) \label{eq:ip_vs_V}
\end{align}

where \textit{g'}, \textit{r'}, and \textit{i'} are the SDSS magnitudes derived from the ILMT observations, and \textit{V} is the apparent magnitude predicted by the MPC based on the object's orbital elements and absolute magnitude (H).
The quality of each fit is quantified by the chi-squared statistic: for the \textit{g'} band, $\chi^2 = 702.21$ with 118 degrees of freedom (DOF) and reduced $\chi^2 \approx 6.0$; for the \textit{r'} band, $\chi^2 = 1293.58$ with 182 DOF (reduced $\chi^2 \approx 7.1$); and for the \textit{i'} band, $\chi^2 = 1034.02$ with 139 DOF (reduced $\chi^2 \approx 7.4$).
Uncertainties on the measured magnitudes include zero-point deviations estimated from neighboring stars, as well as statistical uncertainties including readout noise, Poisson noise, and background noise. Points that deviate significantly from the fitting line correspond to uncertain magnitudes, typically caused by zero point variations or poor PSF fitting. These cases are noted in \ref{notes_tables} and illustrated in \ref{appendix_illustrations} (Figure \ref{fig:asteroid_notes_examples}), which shows representative cutouts of flagged objects. This deviation could also be explained by rotational photometric variability, in which the asteroid’s rotation changes the fraction of sunlight reflected toward the observer, which is not taken into account in the MPC magnitude predictions. When the rotational phase at the time of our observation differs from that of the MPC reference data, this produces additional point-to-point scatter.

The slopes in Equations ~\ref{eq:gp_vs_V}, ~\ref{eq:rp_vs_V}, and ~\ref{eq:ip_vs_V} are close to unity, indicating overall consistency between our photometry and the MPC predictions. However, the reduced $\chi^2$ values are significantly larger than unity in all bands, indicating residual scatter beyond the quoted uncertainties.

To account for this, an additional uncertainty floor was added in quadrature to the individual magnitude uncertainties. The value of this additional term was determined by requiring the reduced $\chi^2$ of the fit to be close to unity, thus accounting for underestimated errors and unmodeled systematics, and was then applied uniformly to all data points. The resulting uncertainty floors are 0.36, 0.32, and 0.35 mag in the \textit{g'}, \textit{r'}, and \textit{i'} bands, respectively.
The resulting color-transformation relations are:
\begin{align}
\textit{g'} &= (0.92 \pm 0.04)\, \textit{V} + (2.05 \pm 0.73) \label{eq:cor_gp_vs_V} \\
\textit{r'} &= (0.97 \pm 0.02)\, \textit{V} + (0.54 \pm 0.49) \label{eq:cor_rp_vs_V} \\
\textit{i'} &= (1.00 \pm 0.03)\, \textit{V} - (0.27 \pm 0.66) \label{eq:cor_ip_vs_V}
\end{align}

The updated fits suggest that the scatter is real and likely reflects systematic effects, such as inaccuracies in the MPC H and G parameters (as noted by \citealt{juric2002comparison}; \citealt{pravec2012absolute}; \citealt{verevs2015absolute}), and color-dependent offsets due to the lack of full transformation corrections.
The fitted intercepts in Equations ~\ref{eq:cor_gp_vs_V}, ~\ref{eq:cor_rp_vs_V}, and ~\ref{eq:cor_ip_vs_V} reveal non-negligible photometric offsets: when both the original and adjusted intercepts are considered, MPC magnitudes tend to be brighter than our SDSS-based estimates by approximately $1.5–2.1$ mag in \textit{g'}, about $0.3 – 0.5$ mag in \textit{r'}, and about $0.3 – 0.4$ mag fainter in \textit{i'}. While part of these offsets may result from the simplified V-band transformation from SDSS filters, which does not fully account for filter differences and asteroid color variations, they also reflect known limitations in the current MPC H and G estimates, as well as possible biases introduced by rotational photometric variability.

\begin{table*}
  \caption{Frequency of observation of the asteroids with the ILMT during its nine commissioning nights from October 24 to November 01, 2022, and during the nights between February 14 and May 16, 2024. The first column indicates the number of nights they were observed. The second and third columns indicate the number of asteroids.}
  \label{tab:freq_table}
  \centering
  \begin{tabular}{c c c c} % centered columns (4 columns)
  
    \hline\hline % inserts double horizontal lines
    {Nights} &
    {Asteroids in Field 0450} &
    {Asteroids in Field 1317}\\ % table heading
    
    \hline % inserts single horizontal line
    1 & 41 & 17 \\
    2 & 33 & 3 \\
    3 & 27 & 1 \\
    4 & 14 & 1 \\
    5 & 7 & 0 \\
    6 & 7 & 0 \\
    7 & 8 & 0 \\
    8 & 8 & 0 \\
    9 & 7 & 0 \\
    \hline % inserts single horizontal line
  \end{tabular}
\end{table*}
%__________________________________________________________________

\subsection{Detection results}
In total, we have made 504 detections of 152 distinct catalogued asteroids in Field 0450, and 30 detections of 22 distinct catalogued asteroids in Field 1317, with predicted \textit{V}-magnitudes ranging from 16.4 to 23.6 and from 17.1 to 21.6, respectively. Table \ref{tab:freq_table} lists the number of asteroids having been observed 1, 2, 3 or more times on different nights. The small number of asteroids observed during several nights is due to their higher average angular speed. The list of the overall detections is provided in \ref{detection_tables}. Many of these asteroids exhibit a complex detection status, particularly in Field 0450: some detections are too faint and approach the detection limit, while others are situated near bright stars that either entirely or partially occult their recorded images. These distinct cases are classified under `Class' in the tables of \ref{detection_tables}, with detailed explanations for the different classes adopted in \ref{appendix_classes}. Zoomed images of asteroids from the different classes are shown in \ref{appendix_illustrations} (Figure \ref{fig:asteroid_classes}). Positions obtained with both methods have an accuracy better than 0.4 arcsec.
%__________________________________________________________________

\subsection{Angular distance calculation}
For each detection, we calculated the angular distance $d$ between the MPC-predicted and the ILMT-observed positions:

\begin{multline}
d = \arccos(\sin(Dec_{obs}) \sin(Dec_{MPC}) \\
 + \cos(Dec_{obs}) \cos(Dec_{MPC}) \cos(\alpha_{MPC} - \alpha_{obs}))
\label{eq:distance}
\end{multline} where $\alpha_{obs}$, $\alpha_{MPC}$, $Dec_{obs}$, $Dec_{MPC}$ are the observed and the MPC-predicted right ascensions, and the observed and MPC-predicted declinations of the asteroid, respectively.
%__________________________________________________________________

\subsection{Astrometric analysis}
\label{astrometric_analysis}
As far as the ILMT observations are concerned, we have determined the UTC at which an asteroid was observed from its right ascension, which exactly corresponds to the local sideral time when the asteroid crossed the central row of the CCD. This UTC is accurate to within several seconds at most.

When using the Minor Planet Checker (MPChecker) to find the predicted position of the same asteroid, we can only specify the observation date with a precision of 0.01 day\footnote{https://www.minorplanetcenter.net/cgi-bin/checkmp.cgi}, i.e., 14.4 minutes. For instance, if the ILMT UTC is 13h 08m 02s on 14 August 2024, the date to be entered in the MPC Checker should be: 2024 08 14.55.
The MPChecker then provides the positions in RA and Dec of the closest asteroids to our measured RA and Dec. The angular separations are calculated using the previously given equation, based on these MPC positions and the ILMT observed positions.  

\begin{figure*}
  \includegraphics[width=17cm, height=.25\textheight]{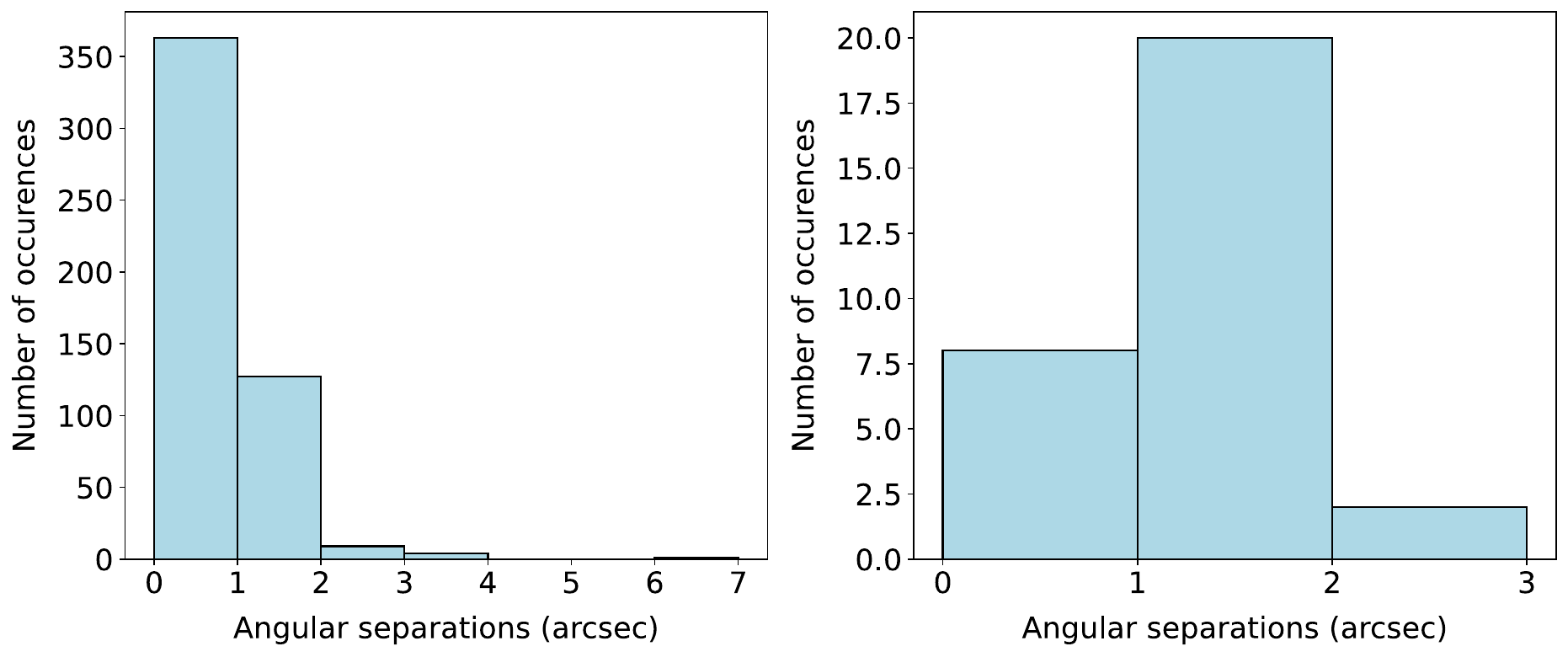}
  \caption{Histogram of residual angular separations for asteroids detected in Field 0450 (left) and Field 1317 (right).}
  \label{fig:remaining_sep_euc_0450_1317}
\end{figure*}

However, the indicated MPChecker epoch corresponds to an MPC UTC of 13h 12m 00s. During the time difference between the ILMT UTC and the quoted MPC UTC, \(\Delta t = 3\,\mathrm{min}\,58\,\mathrm{s}\), the asteroid moved by an angle equal to \(v \times \Delta t\), where \(v\) is the asteroid's angular velocity. The equatorial components of these angular velocities are also provided by the MPChecker.
After correction for these motions, the offsets between the observed ILMT positions and the corrected MPC ones give rise to the corrected histograms illustrated in Figure \ref{fig:remaining_sep_euc_0450_1317} for the Field 0450 and Field 1317.

The measured angular separations between MPC-predicted and ILMT-observed positions of detected asteroids remain limited, ranging from $0''$ to $6.9''$ in Field 0450 and from $0''$ to $6.5''$ in Field 1317. The complete range of angular separations is mentioned in the tables of \ref{detection_tables}.

Numerically, we expect typical differences between MPC-predicted and our observed positions to mostly range between $0''$ and $3.1''$, with possible values reaching up to $5.3''$ in Field 0450. In Field 1317, differences generally range between $0''$ and $6.3''$. These expectations justify our choice of separation thresholds for secure identification.

\begin{figure}
    \resizebox{\hsize}{!}{\includegraphics{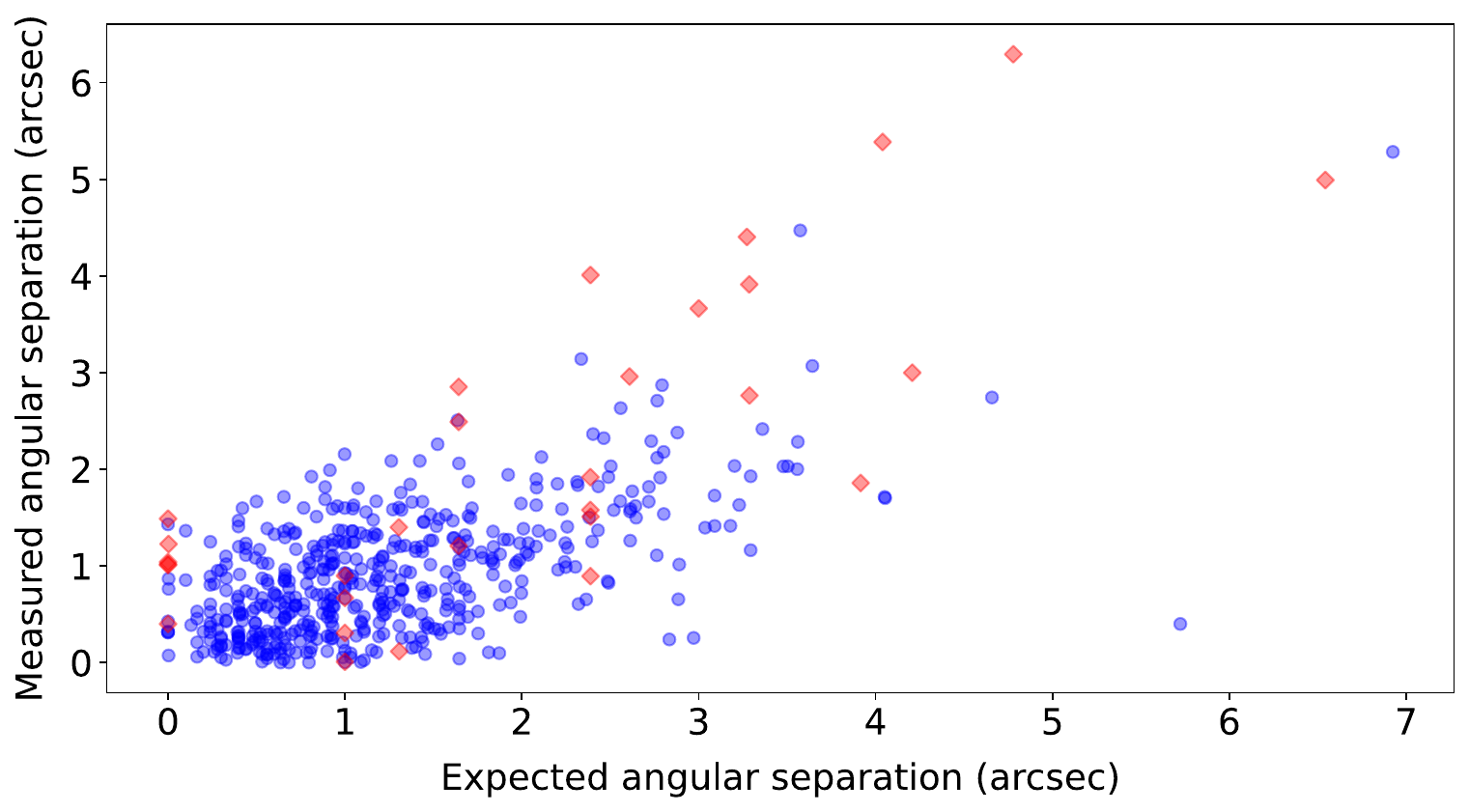}}
    \caption{Scatter plot of expected versus measured angular separations between ILMT-observed and MPC-predicted positions of the detected asteroids. Blue markers represent data from Field 0450 and red markers represent data from Field 1317.}
    \label{fig:measured_vs_th_sep}
\end{figure}

For each detection, we considered the identification secure if the angular separation between the observed and MPC-predicted positions is less than $3''$.  Figure \ref{fig:remaining_sep_euc_0450_1317} shows the histogram of residual angular separations in Fields 0450 and 1317. When the separation exceeds this threshold, we performed a detailed analysis of the equatorial separation components \(\Delta \mathrm{RA} \cdot \cos(\mathrm{Dec}), \ \Delta \mathrm{Dec}\) in relation to the instantaneous velocity components provided by the MPC at the observation time. Specifically, for each detection, we computed the expected positional shift in each coordinate with:

\begin{align}
    \Delta \mathrm{RA}_{\text{pred}} \cdot \cos(\mathrm{Dec}) &= v_{\mathrm{RA}} \cdot \Delta t, \\
    \Delta \mathrm{Dec}_{\text{pred}} &= v_{\mathrm{Dec}} \cdot \Delta t.
\end{align}

where $\Delta RA$ and $\Delta Dec$ are the angular differences between the observed and MPC-predicted positions in equatorial coordinates, the $cos(Dec)$ factor is computed using the MPC-predicted declination, $\Delta t$ is the time offset between the MPC-reported epoch (rounded to 0.01 days) and the precise time of our observation, and $v_{\mathrm{RA}}$ and $v_{\mathrm{Dec}}$ are the equatorial components of angular velocity provided by the MPC, with $v_{\mathrm{RA}}$ already including the $cos(Dec)$ factor. If the difference between observed and theoretical displacements is less than $2''$ in both coordinates, the identification was confirmed;  otherwise, it was classified as uncertain and flagged accordingly in the detection tables. Applying this criterion, only 4 out of 504 detections in Field 0450 were flagged as uncertain, whereas all detections in Field 1317 were confirmed. The correlation between the expected and measured angular separations is strong in both fields, with Pearson coefficients of $0.82$ and $0.79$ for Fields 0450 and 1317, respectively (see Figure \ref{fig:measured_vs_th_sep}).

As shown in this section, the MPC-based cross-checking introduces a time uncertainty that requires the statistical framework presented in this section. This could in principle be avoided by identifying candidate asteroids via MPChecker and retrieving their higher-precision ephemerides directly from NASA's JPL Horizons \citep{giorgini1996jpl}. However, the precision achieved is already sufficient for the scientific objectives of this study. We therefore leave the integration of JPL Horizons ephemerides as a possible refinement for future work, where it could both simplify the cross-matching procedure and enhance its astrometric precision.


%__________________________________________________________________

\subsection{Photometric analysis}

\begin{figure*}
  \centering
  \includegraphics[width=17cm]{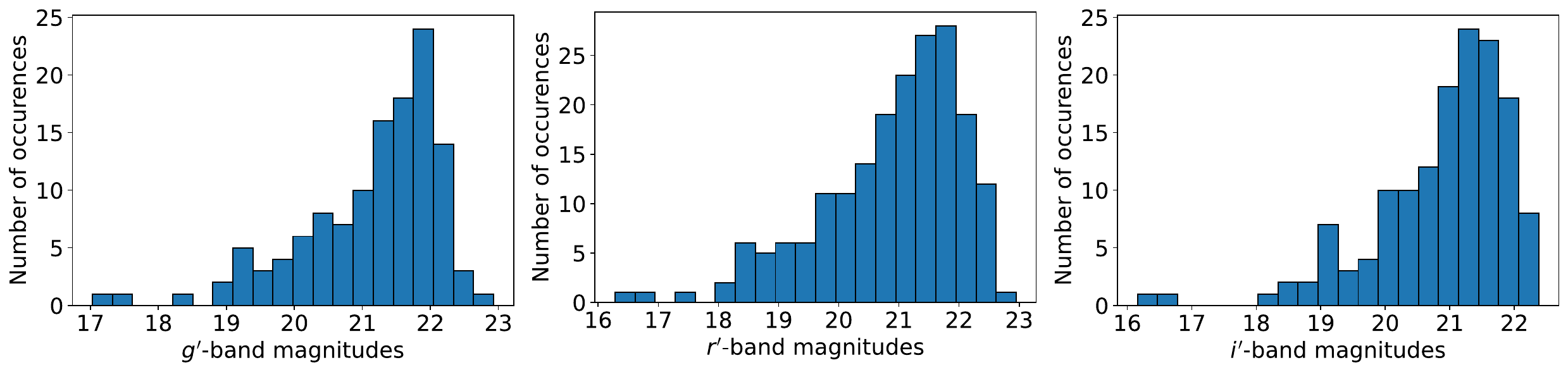}
  \caption{Magnitude distribution of the detected asteroids in Field 0450 through the three SDSS filters.}
  \label{fig:sdss_mag_histo_0450}
    \centering
    \includegraphics[width=17cm]{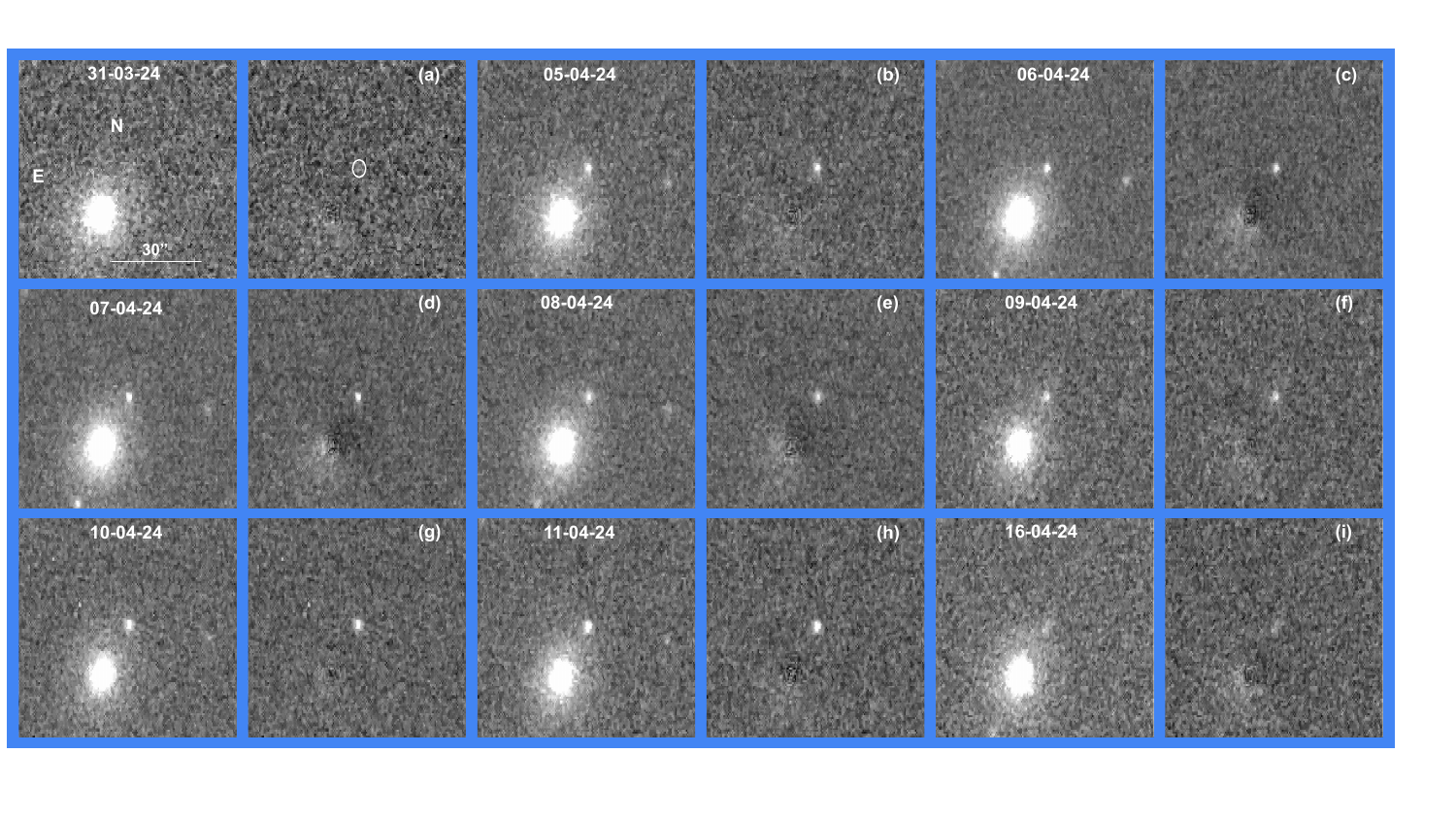}
    \caption{Detection of the transient AT 2024fxn in ILMT images. From left to right and top to bottom order, are nine pairs of images, each leftmost (resp. rightmost) representing the science (resp. subtracted) image, where one can see the transient AT 2024fxn near each frame center. In the same order, observation filters are: \textit{r'}, \textit{r'}, \textit{i'}, \textit{i'}, \textit{i'}, \textit{r'}, \textit{r'}, \textit{r'}, and \textit{i'}.}
    \label{fig:subt_at2024fxn}
\end{figure*}

We have presented in Figure \ref{fig:sdss_mag_histo_0450} the SDSS magnitude histograms of asteroids detected in Field 0450 under very favorable seeing and sky transparency conditions. As shown in that figure, detections reach 22.9, 22.9, and 22.4 in the \textit{g'}, \textit{r'}, and \textit{i'} bands, respectively. This finding is very encouraging.

%__________________________________________________________________

\begin{figure*}
    \centering
    \includegraphics[width=17cm]{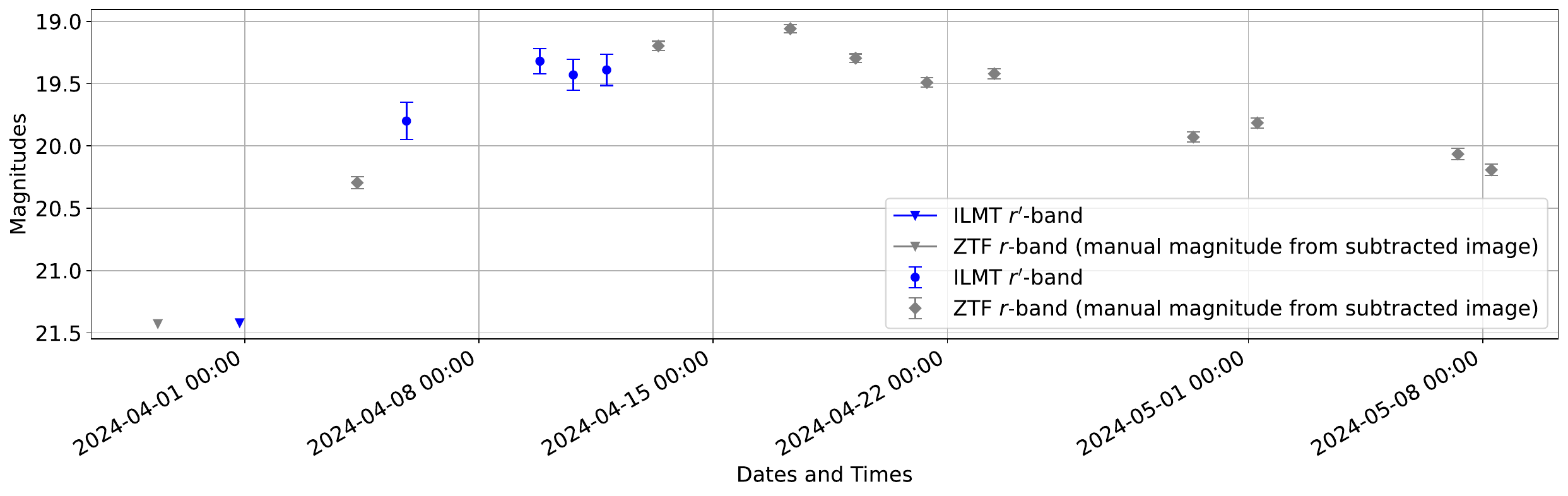}
    \caption{Lightcurve of AT 2024fxn based on ILMT and ZTF difference images. Blue symbols represent magnitudes estimated from the ILMT observations, grey represent \textit{r}-band magnitudes estimated from ZTF observations, triangles represent magnitude estimates with low statistical significance.}
    \label{fig:difference_lightcurve_at2024fxn}
\end{figure*}

\section{Photometric transient: AT 2024fxn}
\label{sec:phot_transient}
\subsection{Methodology}

While running the PyLMT pipeline for Field 1317, we detected several transients that the pipeline classified as belonging to extended hosts. From these, we aimed to select one transient with repeated observations and a noticeable brightness change. We recognised a candidate at the position $\alpha_{2000} = $ 14:13:49.60, $\delta_{2000} = $+29:23:32.0.

This transient has been observed in the \textit{r'} and \textit{i'} bands during 9 different nights with a seeing around $1.8''$, and a first rise in flux observed on April 5th, 2024. The PyLMT classified it as belonging to an extended host with a CNN confidence score of (0.95173323, 0.9997887), representing classification and detection scores, respectively \citep{pranshu2025pylmt}; thus, a possible supernova. We sent an alert to the Transient Name Server (TNS)\footnote{\protect\url{https://www.wis-tns.org/object/2024fxn/discovery-cert}} and the transient has been designated AT 2024fxn \citep{2024TNSTR.964....1P}. In the following hours and days, AT 2024fxn has been reported by ATLAS \citep{tonry2018atlas}, the Zwicky Transient Facility (ZTF, \citealp{bellm2018zwicky}) and Pan-STARRS \citep{magnier2020pan}, with an excellent agreement in position.

We then identified the corresponding science frames containing AT 2024fxn. The \textit{g'}-band provides no usable constraints, as only one underexposed frame was obtained. While several \textit{i'}-band frames are available, their low signal-to-noise due to poor observing conditions results in uncertainties too large for meaningful photometry. Consequently, the photometric analysis is restricted to the \textit{r'}-band. Although the \textit{i'}-band data are not suitable for photometric analysis, image subtraction was nevertheless performed to assess the presence and morphology of the transient; we therefore reconducted a subtraction based on the same reference image per filter, using our custom built ILMTDiff image subtraction software \citep{pranshu2025pylmt} over a square cutout of $2048 \times 2048 $ pixels. For the subtraction to be of ideal quality, reference images should present a better seeing than science images, and a low background level \citep{pranshu2025pylmt}. Therefore, we used specific reference images from 15-02-2024 for the \textit{r'}-band and from 08-02-2024 for the \textit{i'}-band. Figure \ref{fig:subt_at2024fxn} illustrates the results of these subtractions.

We proceeded with the construction of the lightcurve of AT 2024fxn by combining ILMT data with those from the ZTF, noting that the ZTF r-filter profile differs by no more than 0.01 mag from the SDSS \textit{r'} band \citep{suberlak2021improving} across a wide range of g–i colour indices. This approach allowed for enhanced temporal coverage and increased the accuracy of the analysis. Since the transient was heavily contaminated by its neighbouring galaxy, we calculated its magnitudes from subtracted images.

We used the ZTF service\footnote{\myurl}, which provides difference images in the three ZTF spectral filters. In these images, the seeing ranges between $1.8$ and $2.5 \ pixels$ (corresponding to $1.8^{\prime \prime}$ and $2.5^{\prime \prime}$). We calculated the target's magnitudes on the subtracted images using aperture photometry with imexam. Following ZTF recommendations, the optimal photometric aperture radius on these frames is approximately $2 \ pixels$. We tested this choice on another target observed in ZTF frames with similar seeing conditions, for which ZTF also provides a published PSF-photometry light curve. The comparison showed excellent agreement. For our target, we adopted this aperture size and used only \textit{r}-band images, as \textit{i}-band images were limited during this observation period.
Uncertainties in the ILMT magnitudes were estimated by extrapolating the relation between standard deviation and magnitude for ten nearby non-variable reference stars, with the mean zero-point error of these stars added in quadrature.

Uncertainties in the ZTF magnitudes were estimated from two components: (i) the RMS deviations of the magnitude zero points provided in the ZTF difference image products under the MAGZPRMS keyword, and (ii) the statistical component we calculated from the ZTF difference images, which includes Poisson noise, read-out noise, and background noise (added in quadrature).

The zero points for the ILMT magnitudes were estimated from the science images using a highly photometrically stable nearby star, whose stability was verified against five nearby comparison stars. The zero point used in the ZTF magnitude calculations is provided in the ZTF file headers under the MAGZP keyword.

%__________________________________________________________________

\subsection{Results}

The resulting lightcurve is presented in Figure \ref{fig:difference_lightcurve_at2024fxn}. It exhibits a decline of approximately 1 magnitude over the first week, followed by a slower fading over the subsequent two weeks. It is important to highlight that the transient was absent in science images prior to April 4th and after May 8th 2024, in data from both the ILMT and the ZTF. This overall light curve morphology, in addition to the spatial association with a classified galaxy, is partially consistent with certain types of supernovae. However, the data are insufficient to exclude alternative interpretations, such as a foreground classical nova. Additional observations across different filters are necessary in order to constrain the nature of this event.

%__________________________________________________________________

\section{Conclusions}
\label{sec:conclusions}

We have examined CCD frames corresponding to two distinct ILMT fields: Field 0450, starting at $\alpha_{2000} = $ 4h 50m and corresponding to low ecliptic and galactic latitudes, and Field 1317, starting at $\alpha_{2000} = $ 13h 17m and corresponding to high ecliptic and galactic latitudes. In Field 0450, we successfully identified 152 distinct asteroids corresponding to 504 detections in total. In Field 1317, we made 30 detections corresponding to 22 unique known MPC asteroids, with one additional detection being a supernova candidate.

We compared the asteroid detectability in these fields for the different SDSS filters. The magnitudes reached in the SDSS \textit{g'}, \textit{r'}, and \textit{i'} bands under very good seeing and sky transparency conditions are 22.9, 22.9, and 22.4, respectively. The ILMT limiting recall \textit{V}-magnitudes using the PyLMT pipeline are 21.5, 22.0, and 22.0 for observations in these respective bands in Field 0450. For Field 1317, these values are 21 for both \textit{r'} and \textit{i'} bands.

We performed a quantitative comparison between our derived SDSS-band magnitudes and the MPC-predicted V-band magnitudes for asteroids detected in Field 0450 by fitting a linear relation to the data. This comparison is intended as a diagnostic to validate the photometric calibration of our observations, assess systematic offsets between the SDSS and V magnitude systems, and evaluate the reliability of the MPC-predicted brightnesses. Fitted slopes are near unity, confirming general consistency, but with large scatter requiring an uncertainty floor of $\sim$0.32–0.36 mag, depending on the filter. This scatter likely reflects systematic effects from MPC H-magnitude and G-parameter inaccuracies, uncorrected color terms, and rotational phase differences. The fitted intercepts reveal significant photometric offsets between the two datasets: $\sim$1.5–2.1 mag in \textit{g'}, $\sim$0.3–0.5 mag in \textit{r'}, and $\sim$0.3–0.4 mag in \textit{i'}. These results highlight the need for caution when using MPC magnitudes in precise photometry, the application of an uncertainty floor of at least 0.3 mag when comparing with catalog predictions, and further investigation of possible color dependencies and rotational modulation effects. Future work will focus on improved color corrections using SDSS colors, phase-angle adjustments, and cross-comparisons with additional surveys such as Pan-STARRS and Gaia.

We verified the agreement between asteroid positions predicted by the MPC and their counterpart measurements with the ILMT.
Angular separations between observed and predicted positions of the identified asteroids are consistent with expectations when accounting for the distribution of asteroid angular proper motions and the limited time precision of the MPC cross-checking method. Only a small fraction of cases—four out of the 534 asteroids observed in the two investigated fields—show larger discrepancies and require further investigation.

Future calculated positions might be of importance for enhancing the accuracy of trajectories listed in the MPC database. Calculated magnitudes in the three SDSS bands are of a particular interest for further photometric analysis of these asteroids.

SNe candidates are present among the ILMT data, and are worth follow-up identification. We captured a previously uncatalogued photometric transient in Field 1317 and constructed its lightcurve supplementing the ILMT data with ZTF one. The resulting lightcurve is compatible with a SN-hypothesis, however the rise in flux was rather brief and color information is currently unavailable.

These findings reaffirm the ILMT's efficiency as an excellent asteroid hunter, and a promising extragalactic object identifier, aligning with and complementing prior works (see \citealt{pospieszalska2023detection}, \citealt{pranshu2023automated}, \citealt{pranshu2025pylmt}, \citealt{akhunov2024}).
Additionally, dozens more fields from previous observation cycles are yet to be studied. We expect to detect more asteroids and photometric transients through further examination of these fields and improvements of the pipeline.

%__________________________________________________________________

\section*{Acknowlegements}
The 4-m International Liquid Mirror Telescope (ILMT) project results from a collaboration between the Institute of Astrophysics and Geophysics (University of Liège, Belgium), the Universities of British Columbia, Laval, Montreal, Toronto, Victoria and York, and Aryabhatta Research Institute of Observational SciencES (ARIES, India). The authors thank Ankit Bisht, Manisha Kharayat, Hitesh Kumar, Himanshu Rawat, Khushal Singh, Nikhil Dharkiya and other observing staff for their assistance at the 4-m ILMT. The team acknowledges the contributions of ARIES's past and present scientific, engineering and administrative members in the realisation of the ILMT project. JS wishes to thank Service Public Wallonie, F.R.S.-FNRS (Belgium) and the University of Liège, Belgium for funding the construction of the ILMT. PH acknowledges financial support from the Natural Sciences and Engineering Research Council of Canada, RGPIN-2019-04369. SF, PH, AP-S, and JS thank ARIES for hospitality during their visits to Devasthal. KM, BK, BA and ND acknowledge the support from the BRICS grant DST/ICD/BRICS/Call-5/CoNMuTraMO/2023 (G) funded by the DST, India. M.D. acknowledges Innovation in Science Pursuit for Inspired Research (INSPIRE) fellowship award (DST/INSPIRE Fellowship/2020/IF200251) for this work. T.A. thanks Ministry of Higher Education, Science and Innovations of Uzbekistan (grant FZ-20200929344). JS and KM acknowledge the assistance received from the Anusandhan National Research Foundation (ANRF, SERB- 762 VAJRA Faculty Scheme, India).

This work makes use of data and/or services provided by the International Astronomical Union's Minor Planet Center.

\textit{Software}: Astropy (Astropy Collaboration et al. 2013, 2018, 2022), Matplotlib \citep{hunter2007matplotlib}, NumPy \citep{harris2020array}, imexam \citep{sosey2017imexam}, SAOImage DS9 \citep{joye2003new}.
\vspace{-1em}

%__________________________________________________________________

\bibpunct{(}{)}{;}{a}{}{,}
\bibliography{refs_new} % your references Yourfile.bib

%__________________________________________________________________

%%Appendix

\onecolumn
\appendix
\renewcommand{\thesection}{Appendix \Alph{section}}

%__________________________________________________________________

\section{Detection classes}
\label{appendix_classes}
We classified the detected asteroids according to different detection classes:

\begin{itemize}
\item{MP: The asteroid appears in the MPC, and was detected by the CNN algorithm.}
\item{MN: The asteroid appears in the MPC, and was detected visually.}
\item{MNBS: The asteroid appears in the MPC, was detected visually near a Bright Star, and was not detected by the CNN algorithm due to the angular proximity to that Bright Star.}
\item{MNF: The asteroid appears in the MPC, and was detected visually, but is Faint.}
\item{MNVF: The asteroid appears in the MPC, and was detected visually, but is Very Faint.}
\item{MN-MP: The asteroid was detected by the CNN algorithm after tuning its parameters.}
\end{itemize}
%__________________________________________________________________

\section{Photometry parameters}
\label{appen:phot_params}
Asteroid magnitudes are estimated through aperture photometry. The catalogues and band-passes used for the calibrators are:
\begin{itemize}
\item\textit{
For the \textit{r'}-band detections: The Guide Star Catalog, Version 2.4.2 (GSC2.4.2) (STScI, 2020), rmag values (SDSS r-band).}

\item\textit{
For the \textit{g'}-band detections: The Guide Star Catalog, Version 2.4.2 (GSC2.4.2) (STScI, 2020), gmag values (SDSS g-band).}

\item\textit{
For the \textit{i'}-band detections: The Guide Star Catalog, Version 2.4.2 (GSC2.4.2) (STScI, 2020), imag values (SDSS i-band).}
\end{itemize}

Note: The magnitude uncertainties are estimated based upon the zero point variations and are intended to reflect the stability of the photometric calibration. Generally, this zero point uncertainty is smaller than 0.1 mag; otherwise, a note is provided in the table's rightmost column. Unless specified otherwise, a minimum of three neighbouring stars has been used to estimate the zero point. The full photometric uncertainty budget is described in Section \ref{subsec:mag_and_pos}. Measurements flagged with `:' or `::' indicate high photometric uncertainty and are excluded from the analysis described in Section \ref{subsec:mag_and_pos}.
%__________________________________________________________________

\section{Asteroid detection tables}
\label{detection_tables}
\subsection*{Description of the tables}

Tables \ref{tab:table_0450} and \ref{tab:field_1317} present the asteroids found in Field 0450 and at $\alpha=$ 13h 17m, respectively. The different targets were sorted following the order of their increasing right ascension.\\

Table \ref{tab:table_0450} contains: date and time of observation, J2000.0 equatorial coordinates predicted by the MPC, angular distance between MPC-predicted and ILMT-observed positions, MPC-predicted V magnitude, MPC name, ILMT detection class, filter used for ILMT observation (F), J2000.0 equatorial coordinates observed by the ILMT, ILMT photometry in the given filter, type of photometry (Flag), and Notes.\\
The cited filters are from the SDSS. `Flag' specifies the method used to calculate the magnitude and/or position: 
`pho1' refers to the \textit{statistical} method, `pho2' to the \textit{individualised} method, and `pho3' to a position estimate derived from the CNN algorithm, based on visual examination rather than centroid fitting.
The `Notes' column provides further information about the aperture photometry parameters and is further developped in \ref{notes_tables}.\\

In Table \ref{tab:field_1317} we present the same columns as in Table \ref{tab:table_0450}, with the exception of the `Notes' column.\\

% Reset table counter and customize table numbering format
\refstepcounter{table}
\captionsetup{justification=centering} 
\label{tab:table_0450} % Label for referencing
\includepdf[pages=-, angle=90]{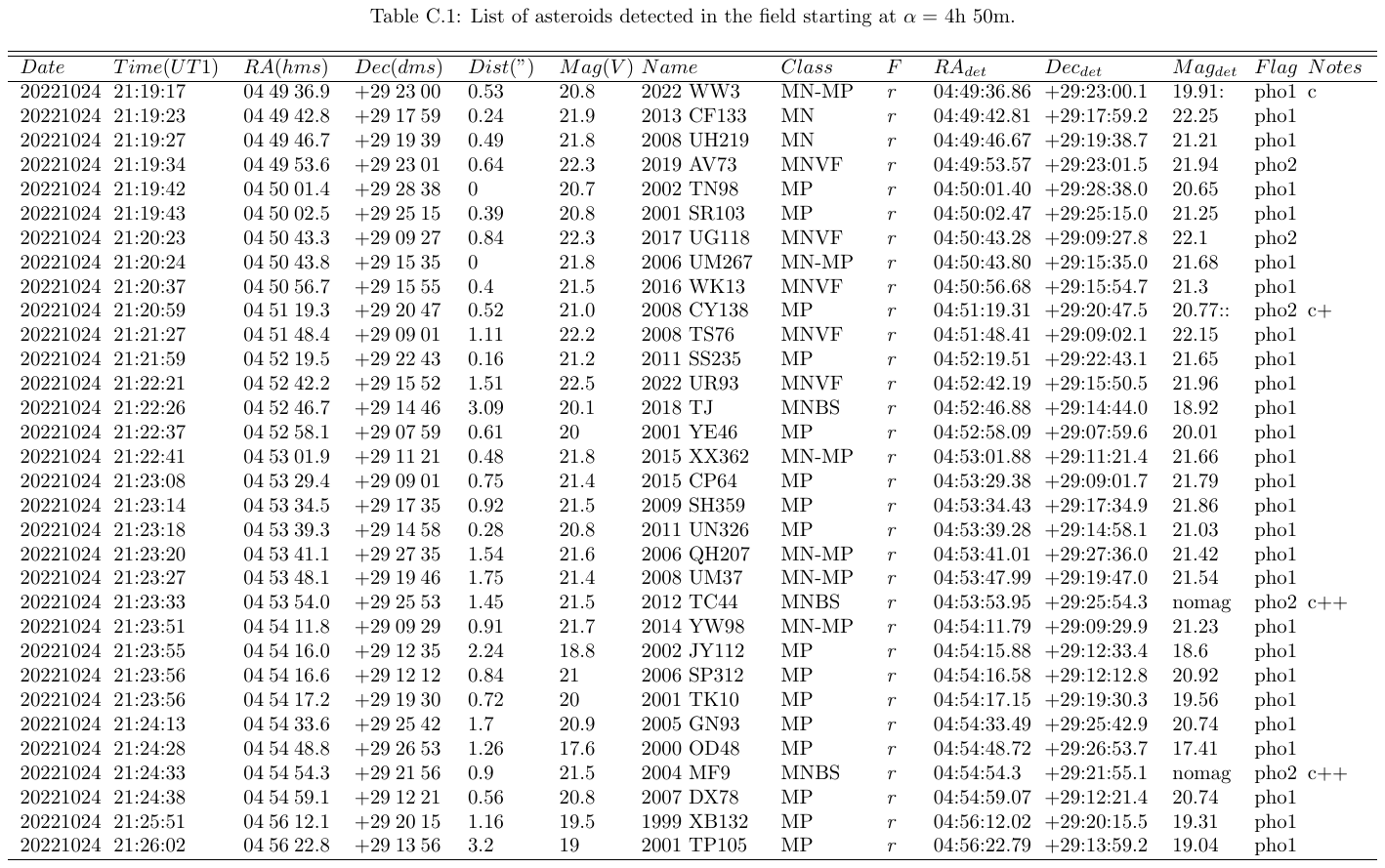} % Insert the PDF, where "pages=-" includes all pages

%\addtocounter{table}{1}
\begin{table*}
  \captionsetup{width=\textwidth, labelformat=empty} % Adjust width and remove numbering
  \caption{Table C.2: List of all MPC detections in Field 1317.\label{tab:field_1317}}
  \small
  \centering
  \begin{tabular}{p{1.1cm}p{1.3cm}p{1.3cm}p{1.3cm}p{0.8cm}p{0.7cm}p{1.9cm}p{0.3cm}p{0.2cm}p{1.2cm}p{1.0cm}p{0.3cm}p{0.3cm}}
  
    \hline\hline % inserts double horizontal lines
    
    \raggedright $Date$& {$Time(UT1)$}& \multicolumn{1}{c}{$RA(hms)$}& $Dec(dms)$& {\raggedleft {$Dist('')$}}& {$Mag(V)$}& \multicolumn{1}{c}{$Name$}& \multicolumn{1}{l}{\raggedleft $Class$}& $F$& $RA_{MPC}$& $Dec_{MPC}$& \multicolumn{1}{l}{\raggedleft$Mag_{det}$}& $Flag$\\
    
    \hline % inserts single horizontal line
    20240217&	22:21:30&	13:28:42.0& +29:30:58& 2.39& 20.8& 2000 WR124  & MP& \textit{i'}&	13:28:41.9& +29:31:00&	19.80& pho2\\
    20240217&	22:24:26&	13:31:39.2& +29:34:12& 2.39& 19.7& 2002 JJ58   & MP& \textit{i'}&	13:31:39.1& +29:34:14&	19.14& pho2\\
    20240217&	22:26:15&	13:33:29.1& +29:37:15& 4.21& 19.6& 1999 PM1    & MN& \textit{i'}&	13:33:29.0& +29:37:19&	18.97& pho2\\
    20240217&	22:29:22&	13:36:37.0& +29:29:39& 3.00& 19.2& 2000 FK1    & MP& \textit{i'}&	13:36:37.0& +29:29:36&	19.23& pho2\\
    20240217&	22:30:44&	13:37:59.1& +29:19:04& 0.00& 21.4& 2014 WF498  & MNF& \textit{i'}&	13:37:59.1& +29:19:04&	20.51& pho2\\
    20240217&	22:32:17&	13:39:32.0& +29:36:14& 0.00& 19.6& 2002 FM6    & MP& \textit{i'}&	13:39:32.0& +29:36:14&	19.18& pho2\\
    20240217&	22:39:19&	13:46:35.0& +29:35:48& 6.54& 19.8& 2006 CV56   & MP& \textit{i'}&	13:46:35.2& +29:35:54&	19.74& pho2\\
    20240217&	23:29:18&	14:36:46.9& +29:29:21& 0.00& 20.8& 2009 XS3    & MP& \textit{i'}&	14:36:46.9& +29:29:21&	20.31& pho2\\
    20240331&	19:57:20&	13:53:42.7& +29:39:39& 1.64& 19.1& Martinphillipps& MP& \textit{r'}& 13:53:42.6& +29:39:41&	18.78& pho2\\
    20240331&	20:09:01&	14:05:26.3& +29:36:27& 1.00& 19.4& 2012 BJ24   & MP& \textit{r'}&	14:05:26.3& +29:36:28&	19.33& pho2\\
    20240331&	20:20:32&	14:16:58.9& +29:36:55& 1.64& 19.6& 2000 WD151  & MP& \textit{r'}&	14:16:59.0& +29:36:54&	19.66& pho2\\
    20240405&	19:48:37&	14:04:40.6& +29:21:50& 4.78& 20.5& 2002 QB80   & MP& \textit{r'}&	14:04:40.8& +29:21:46&	19.64& pho2\\
    20240405&	19:53:54&	14:09:58.8& +29:35:41& 0.00& 19.5& 2000 WD151  & MP& \textit{r'}&	14:09:58.8& +29:35:41&	19.25& pho2\\
    20240405&	19:58:24&	14:14:29.8& +29:23:17& 2.39& 20.8& 2013 CT129  & MN& \textit{r'}&	14:14:29.7& +29:23:19&	21.76& pho2\\
    20240406&	19:48:28&	14:08:28.1& +29:22:18& 3.27& 19.6& 2003 FC42   & MP& \textit{i'}&	14:08:28.2& +29:22:15&	19.12& pho2\\
    20240406&	19:48:30&	14:08:30.4& +29:33:53& 4.04& 19.5& 2000 WD151  & MP& \textit{i'}&	14:08:30.7& +29:33:54&	18.97& pho2\\
    20240406&	19:53:47&	14:13:49.3& +29:30:57& 1.00& 20.8& 2013 CT129  & MN& \textit{i'}&	14:13:49.3& +29:30:56&	19.87& pho2\\
    20240406&	19:58:11&	14:18:14.8& +29:35:47& 3.29& 21.6& 2007 AS12   & MN& \textit{i'}&	14:18:14.6& +29:35:49&	20.67& pho2\\
    20240407&	19:43:06&	14:07:01.8& +29:31:34& 3.92& 19.5& 2000 WD151  & MP& \textit{i'}&	14:07:01.5& +29:31:34&	19.45& pho2\\
    20240407&	19:43:48&	14:07:43.4& +29:34:33& 2.39& 19.6& 2003 FC42   & MP& \textit{i'}&	14:07:43.3& +29:34:35&	19.09& pho2\\
    20240407&	19:49:11&	14:13:08.0& +29:38:20& 1.64& 20.8& 2013 CT129  & MN& \textit{i'}&	14:13:08.1& +29:38:19&	19.77& pho2\\
    20240410&	19:38:49&	14:14:34.9& +29:18:27& 0.00& 18.8& 1998 QB3    & MP& \textit{r'}&	14:14:34.9& +29:18:27&	18.53& pho2\\
    20240411&	19:34:10&	14:13:51.4& +29:29:54& 2.39& 18.8& 1998 QB3    & MP& \textit{r'}&	14:13:51.5& +29:29:52&	18.73& pho2\\
    20240411&	19:38:52&	14:18:35.6& +29:21:31& 1.00& 19.5& 2005 YN155  & MP& \textit{r'}&	14:18:35.6& +29:21:30&	19.19& pho2\\
    20240416&	18:57:35&	13:56:52.3& +29:34:53& 1.00& 21.1& 2003 AB83   & MNVF& \textit{i'}&	13:56:52.3& +29:34:52&	19.72& pho2\\
    20240416&	19:10:15&	14:09:34.1& +29:23:17& 0.00& 19.8& 2004 NL16   & MP& \textit{i'}&	14:09:34.1& +29:23:17&	19.09& pho2\\
    20240416&	19:14:59&	14:14:19.9& +29:32:51& 1.30& 19.5& 2005 YN155  & MP& \textit{i'}&	14:14:19.8& +29:32:51&	18.72& pho2\\
    20240504&	18:22:20&	14:32:30.7& +29:34:44& 2.61& 18.0& 2001 CE34   & MP& \textit{r'}&	14:32:30.9& +29:34:44&	17.83& pho2\\
    20240512&	17:53:18&	14:34:57.1& +29:38:37& 3.29& 17.1& Hickson     & MP& \textit{r'}&	14:34:57.3& +29:38:39&	15.86& pho2\\
    20240516&	17:31:26&	14:28:47.9& +29:25:02& 1.31& 18.9& 1999 YH9    & MP& \textit{i'}&	14:28:47.8& +29:25:02&	17.96& pho2\\
    \hline % inserts single horizontal line
  \end{tabular}
\end{table*}

%__________________________________________________________________
\section{Illustration of asteroid classes and notes}
\label{appendix_illustrations}

\begin{figure}[H]
    \centering

    \begin{subfigure}[t]{0.3\textwidth}
        \includegraphics[width=\linewidth]{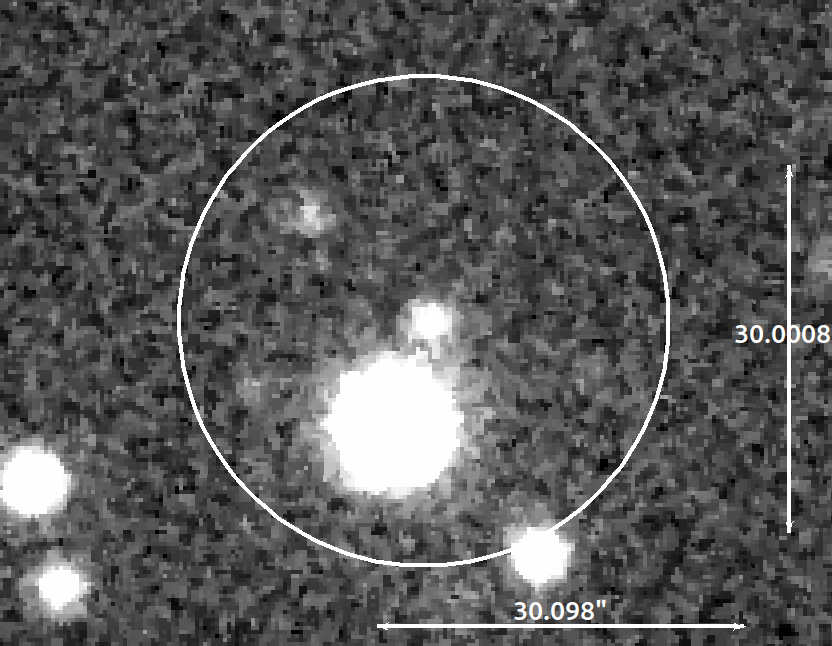}
        \caption*{\textbf{c} \\ Slight contamination by neighbor}
    \end{subfigure}
    \hfill
    \begin{subfigure}[t]{0.3\textwidth}
        \includegraphics[width=\linewidth]{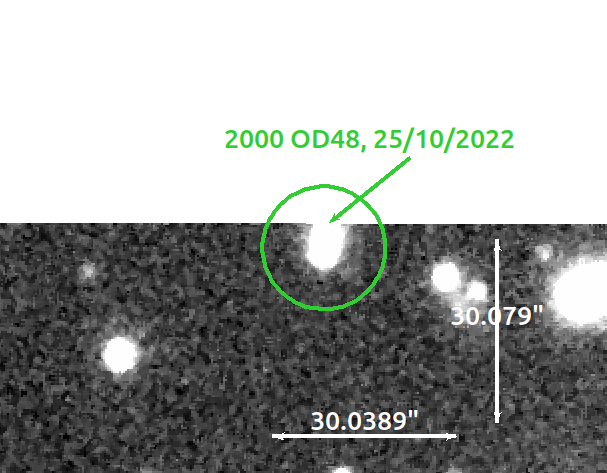}
        \caption*{\textbf{c++} \\ Subtraction only; no magnitude}
    \end{subfigure}
    \hfill
    \begin{subfigure}[t]{0.3\textwidth}
        \includegraphics[width=\linewidth]{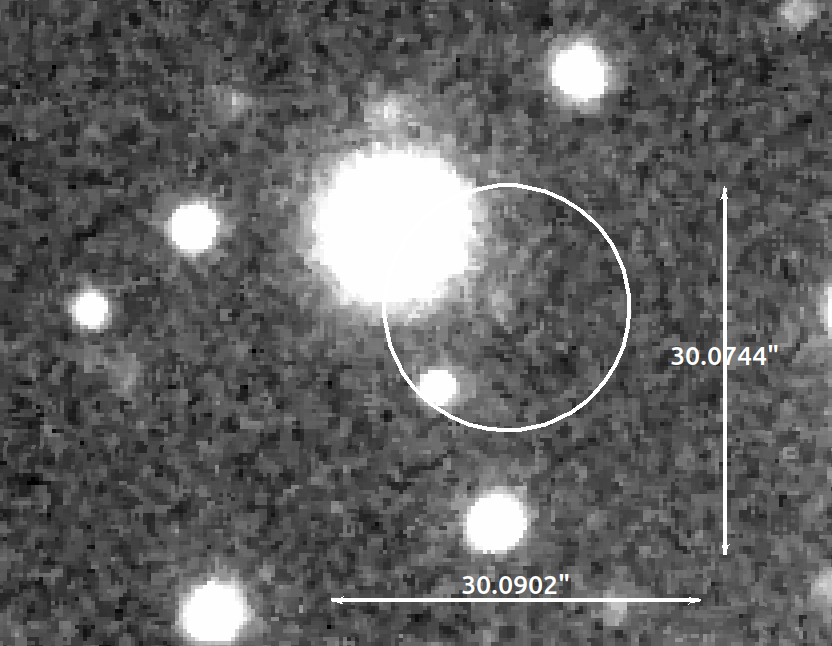}
        \caption*{\textbf{hal} \\ In halo of a bright star}
    \end{subfigure}
    \hfill

    \vspace{1em}
    \begin{subfigure}[t]{0.3\textwidth}
        \includegraphics[width=\linewidth]{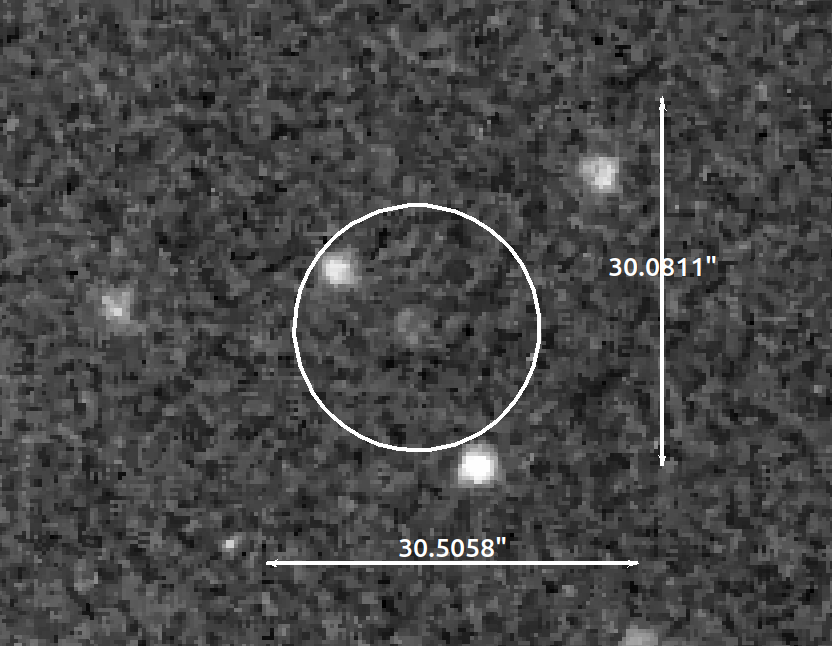}
        \caption*{\textbf{ld} \\ Limit of detection, poor fit}
    \end{subfigure}
    \hfill
    \begin{subfigure}[t]{0.3\textwidth}
        \includegraphics[width=\linewidth]{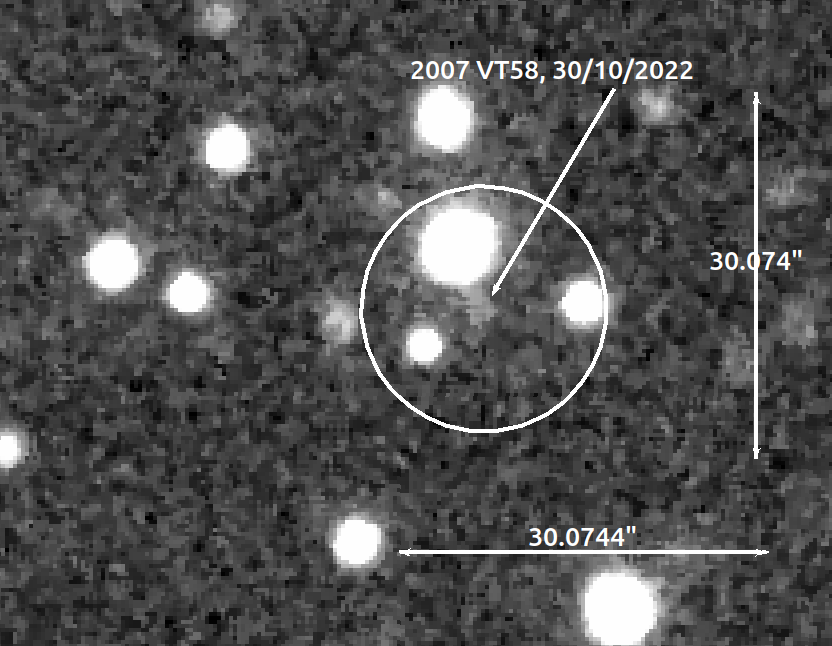}
        \caption*{\textbf{byh} \\ Position estimated visually}
    \end{subfigure}
    \hfill
    \begin{subfigure}[t]{0.3\textwidth}
        \includegraphics[width=\linewidth]{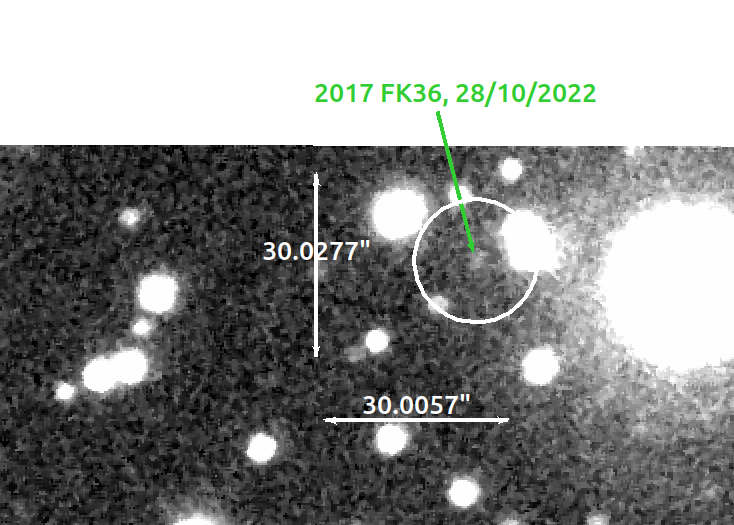}
        \caption*{\textbf{ext} \\ Extended source}
    \end{subfigure}
    \hfill

    \caption{Representative zoomed-in cutouts for objects flagged with specific notes. These include cases of contamination, morphology, or difficulty in measurement. Labels refer to: \textbf{c} — slight contamination; \textbf{c++} — detected only in subtraction; \textbf{hal} — in halo of bright star; \textbf{ld} — near limit of detection; \textbf{byh} — position visually estimated; \textbf{ext} — extended morphology. North is up and East is to the left.}
    \label{fig:asteroid_notes_examples}
\end{figure}

\begin{figure}[H]
    \centering

    \begin{subfigure}[t]{0.3\textwidth}
        \includegraphics[width=\linewidth]{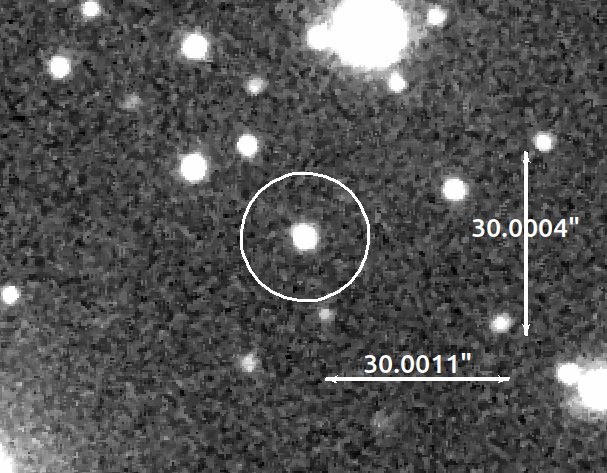}
        \caption*{\textbf{MP} \newline CNN detection (MPC match)}
    \end{subfigure}
    \hfill
    \begin{subfigure}[t]{0.3\textwidth}
        \includegraphics[width=\linewidth]{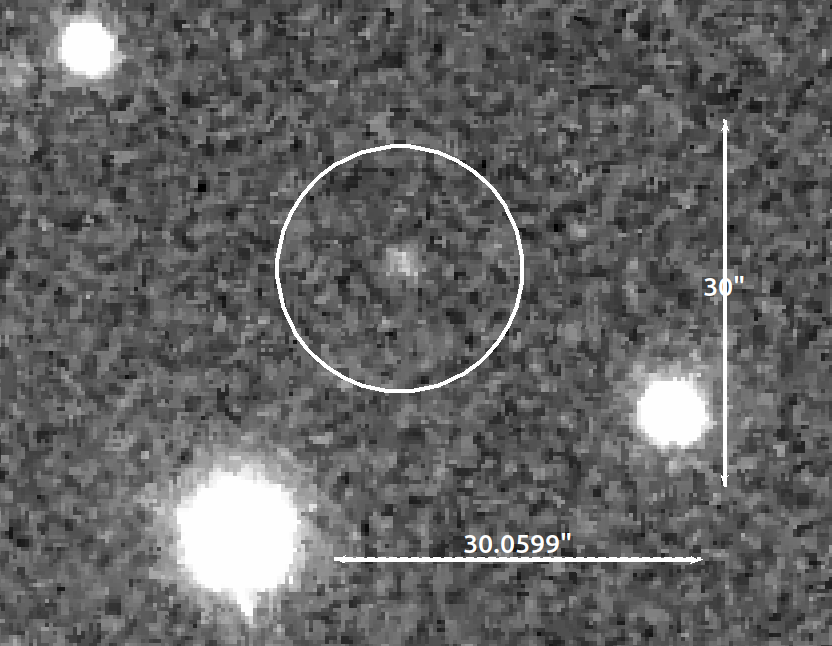}
        \caption*{\textbf{MN-MP} \newline Detected by CNN after tuning}
    \end{subfigure}
    \hfill
    \begin{subfigure}[t]{0.3\textwidth}
        \includegraphics[width=\linewidth]{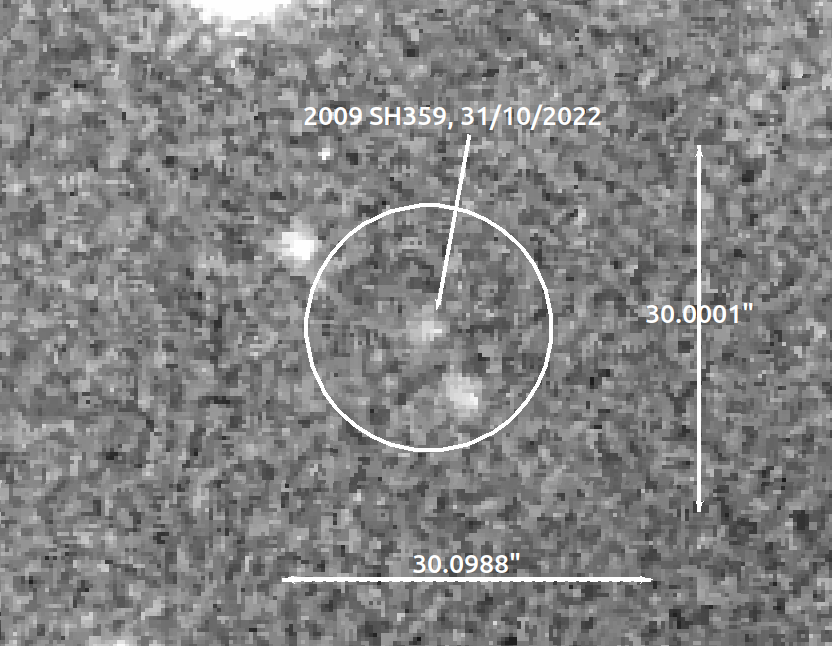}
        \caption*{\textbf{MNF} \newline Visually detected, faint}
    \end{subfigure}

    \vspace{1em}
    \begin{subfigure}[t]{0.3\textwidth}
        \includegraphics[width=\linewidth]{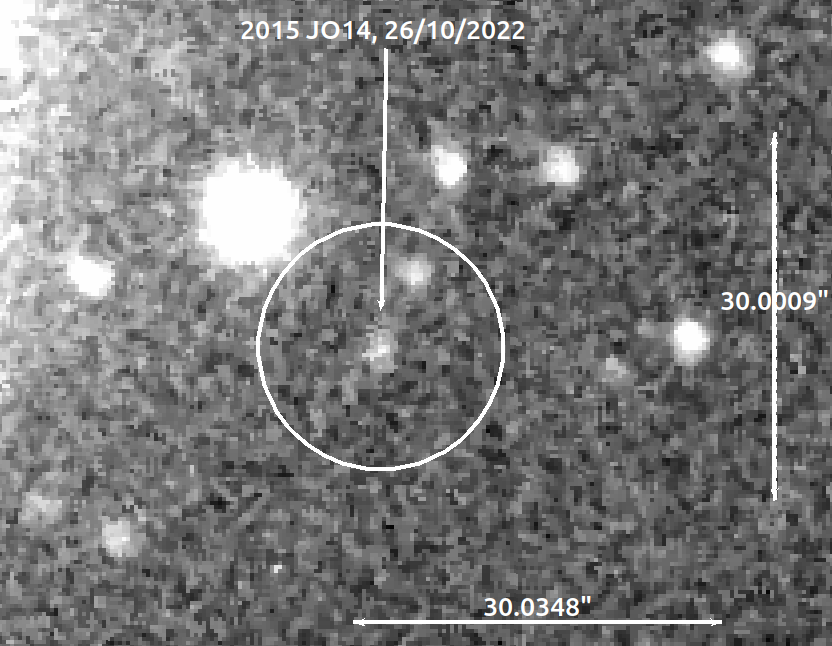}
        \caption*{\textbf{MNVF} \newline Visually detected, very faint}
    \end{subfigure}
    \hfill
    \begin{subfigure}[t]{0.3\textwidth}
        \includegraphics[width=\linewidth]{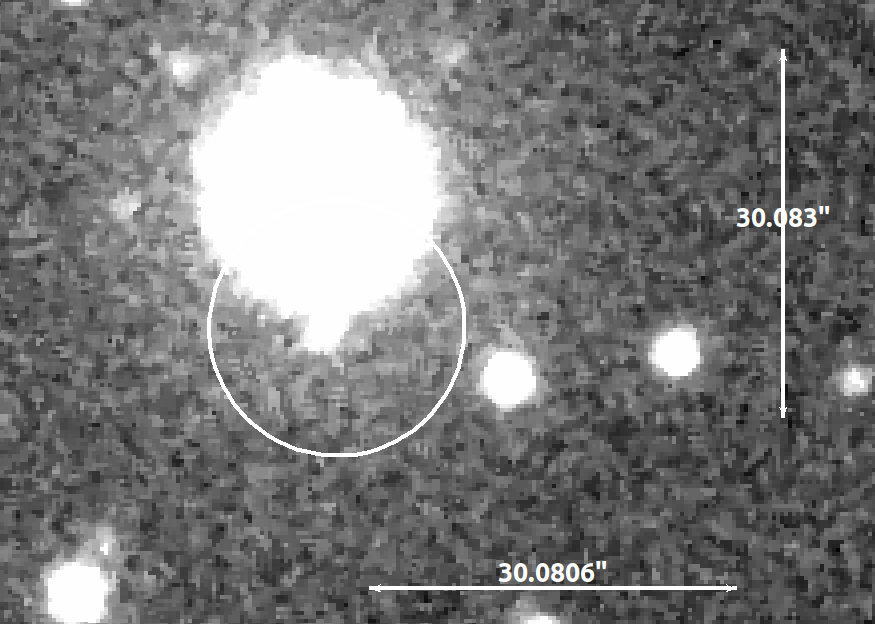}
        \caption*{\textbf{MNBS} \newline Missed by CNN due to bright star}
    \end{subfigure}   
    \hfill
    \begin{subfigure}[t]{0.3\textwidth}
        \includegraphics[width=\linewidth]{ext-MN-2017-20221028FK36.png}
        \caption*{\textbf{MN} \newline Visual detection (MPC match)}
    \end{subfigure}

    \caption{Zoomed-in cutouts of representative asteroids for each classification. The circle markers indicate asteroid positions. North is up and East is to the left.}
    \label{fig:asteroid_classes}
\end{figure}

%__________________________________________________________________

\section{Explanation of the `Notes' column of Table \ref{tab:table_0450}}
\label{notes_tables}
\begin{itemize}
\item{c: Slight contamination due to a neighbour star; an estimated value for the magnitude is given.}
\item{c+: Due to contamination from a neighbouring star, the magnitude could not be calculated with sufficient precision.}
\item{c++: Detected by subtraction, magnitude cannot be calculated.}
\item{hal: Target is in the halo of a very bright star.}
\item{ld: Limit of detection, detected visually, a gaussian profile does not fit perfectly.}
\item{prev: The same reference stars were used as for the previous target.}
\item{2*: No more than two stars were used as references for the zero point calculation.}
\item{byh: The position was estimated visually due to noise or due to a poor gaussian fit.}
\item{ext: The source appears extended, uncertainty on position might follow ($\sim$$1''$).}
\item{2ob: The target appears like two objects instead of one. But it is uncertain due to noise.}
\item{nomag: A magnitude value is not available.}
\item{uncert: Detection considered uncertain or requiring further investigation due to the angular separation, or its components, between the ILMT-observed and MPC-predicted positions exceeding one or more of the cited thresholds.}

\end{itemize}

% End of appendix

\end{document}